\documentclass[useAMS,usenatbib]{mn2e}
\usepackage{graphicx}
\usepackage{subfigure}
\usepackage{natbib}
\usepackage[english]{babel}
\usepackage{amssymb}
\usepackage{amsmath}
\usepackage{xspace}
\usepackage{color}

\newcommand\ionm[2]{#1$\,${\small\rmfamily{#2}}}% 
\newcommand{\hi}{\ionm{H}{I}\xspace}
\def\h2{\ifmmode {\mbox H$_2$}\else H$_2$\fi\xspace}
\def\cosmos{\ifmmode {\mbox {\it COSMOS}}\else {\it COSMOS} \fi}

\title [ISM properties at high-z]{The nature of the ISM in galaxies during the star-formation activity peak of the Universe}
\author[G. Popping, J.P. P\'erez-Beaupuits, M. Spaans, S.C. Trager and R.S. Somerville]{G. Popping$^{1}$\thanks{E-mail: g.popping@astro.rug.nl},
  J.P. P\'erez-Beaupuits$^{2}$, M. Spaans$^{1}$, S.C. Trager$^{1}$ and R.S. Somerville$^{3}$\\
$^{1}$Kapteyn Astronomical Institute, University of Groningen, Postbus 800, NL-9700 AV Groningen, the Netherlands\\
$^{2}$Max-Planck-Institut f\"ur Radioastronomie, Auf dem H\"ugel 69, 53121, Bonn, Germany\\
$^{3}$Department of Physics and Astronomy, Rutgers University, 136
Frelinghuysen Road, Piscataway, NJ 08854, USA}
\begin{document}

\maketitle

\begin{abstract}
We combine a semi-analytic model of galaxy formation, tracking atomic and
molecular phases of cold gas, with a three-dimensional
radiative-transfer and line tracing code to study the sub-mm emission from atomic and molecular species
(CO, HCN, [C I], [CII], [OI]) in galaxies. We compare the physics that
drives the formation of stars at the epoch of peak star formation (SF) in the
Universe ($z = 2.0$) with that in local galaxies. We find that normal star-forming galaxies
at high redshift have much higher CO-excitation peaks than their local counterparts and
that CO cooling takes place at higher excitation levels. CO line ratios increase with
redshift as a function of galaxy star-formation rate, but are well
correlated with \h2 surface density independent of redshift.
We find an increase in the [OI]/[CII] line ratio in typical star-forming
galaxies at $z = 1.2$ and $z = 2.0$ with respect to counterparts at $z = 0$. Our model
results suggest that typical star-forming galaxies at high redshift consist of much
denser and warmer star-forming clouds than their local counterparts. Galaxies belonging to
the tail of the SF activity peak at $z = 1.2$ are already less dense and cooler than
counterparts during the actual peak of SF activity ($z = 2.0$). We use our results to
discuss how future ALMA surveys can best confront our predictions and constrain models of galaxy
formation.
\end{abstract}


\section{Introduction}
The star-formation rate density (SFRD) of the Universe peaked at
redshifts $z\sim 1-3$ after which it gradually dropped towards its
present-day value \citep[e.g.,][]{Hopkins2006}.  This epoch marks a crucial period in the history of the Universe, 
when the bulk of stars in massive galaxies were likely formed.

The development of large samples of galaxies with extensive
multi-wavelength information has been the first step in understanding
the physical conditions under which these galaxies formed their stars. These surveys probe fundamental properties such as
the stellar mass, star-formation rate (SFR) and sizes of these
galaxies \citep[e.g., {\it GOODS}, \cosmos and {\it
  CANDELS}][]{Giavalisco2004,Scoville2007,Grogin2011,Koekemoer2011}. Information about the gas properties of these
objects has largely been unavailable. Direct observations of the gaseous
content of typical star-forming galaxies are only available for a
few dozen objects \citep[e.g.,][]{Daddi2010,Genzel2010,
  Tacconi2010,Geach2011,Bauermeister2013, Tacconi2013}. These
observations are usually limited to measures of the gaseous mass of
galaxies, and do not probe the physical state under which the stars
are formed: the local radiation field, temperature, and density of the
gas are unknown. Some authors have addressed these characteristic
properties of molecular \citep{Dannerbauer2009,Danielson2011} and
atomic gas \citep{Walter2011}, but such studies are limited to the
most actively star-forming or submillimeter galaxies (SMGs), and are
not representative of the bulk galaxy population
\citep{Walter2011}. We refer the reader to \citet{Carilli2013} for a
full overview of observations of the interstellar medium (ISM) in
high-z galaxies.

The temperature, density and turbulence of the gas and the local
radiation field are the main characteristics in play during the
formation of stars out of molecular clouds, as they set the free-fall
time, pressure and sound speed of the gas. Furthermore, the gas
contains several coolant species, responsible for the dominant cooling
of the interstellar medium. In order to fully understand the physics
that drives the bulk of the SF during the peak in SFRD of the
Universe, it is important to probe these gas properties in typical
star-forming galaxies (i.e., responsible for the bulk of SF during the
SF peak of the Universe) and to understand under which conditions
stars formed. It is unclear if the physical state of the gas in these
galaxies is similar to local counterparts or if the ISM physics that drives the SF
efficiency is significantly different. A proper comparison of the
density, pressure, temperature and radiation field of the star-forming
gas in galaxies at $z\sim2.0$ with local galaxies can shed light on
the conditions under which galaxies formed their stars during the SF
peak of the Universe with respect to the local Universe.

In the near future, millimeter telescopes such as ALMA (Atacama Large
Millimeter Array), PdBI (Plateau du Bureau Interferometer), and LMT (Large
Millimeter Telescope) will be able
to observe multiple line transitions of tracers of molecular gas and
atomic gas for statistical samples of high-redshift galaxies. These
observations will not only reveal the physics under which star
formation took place, but will also provide theorists with new
constraints on the physical state of the star-forming cold gas.  It is
therefore important to develop computationally efficient methods to
study the sub-mm line properties of the cold gas in cosmologically
representative galaxy samples.

The chemical evolution of molecular species can be obtained by the modeling of Photon (UV) Dominated Regions (PDR) and X-ray dominated
regions (XDR) \citep[e.g.][]{Meijerink2006,Meijerink2007}. Besides UV
and X-ray heating of the gas, cosmic rays and shocks can also add radiative or mechanical energy to the system, heat the gas
and play a major role in the PDR and XDR chemistry \citep{Loenen2008}. Furthermore, radiative feedback on molecular clouds can also result in a PDR
dominated molecular chemistry. Large velocity gradient (LVG) radiative
transfer codes \citep[e.g.,][]{Weiss2005,Poelman2006,Radex} provide
accurate estimates of line emission along the energy ladder of
molecular and atomic species within PDRs. The LVG formalism permits the treatment of coupled
radiative transfer and molecular excitation as a local problem. It
assumes that the photons emitted from one region do not interact with
molecules in other regions due to Doppler shift. Photons only interact
with molecules in the local region where they were emitted.  These methods have proven to be excellent theoretical
tools to reproduce the sub-mm line emission of various chemical species in
molecular clouds, ranging over a large variety of densities and
temperatures. They do not necessarily encompass the large variety of
gaseous phases in a galaxy's ISM, and the LVG assumption that photons
do not interact with molecules in other regions may not necessarily be
true. A proper comparison between theory and observations of galaxies
therefore requires PDR and radiative transfer methods to be embedded
within cosmological models of galaxy formation and evolution. It
furthermore requires one to properly calculate the interaction of photons
(escape probability) throughout all the grid cells along the line of
sight, taking into account the opacities at each velocity (or frequency) resolution element.

In the last decade large efforts have been made to develop numerical
models in a cosmological context to understand the relation between
dense and cold gas properties and other galaxy properties
\citep[e.g.,][]{Pelupessy2006,Narayanan2008,Robertson2008,Gnedin2009,Wada2009,Kuhlen2012,Christensen2012}. Only
occasionally have these efforts been combined with detailed modeling of
the sub-mm line intensity coming from chemical species such as 
CO, [CII] and HCN \citep{Narayanan2008,JP2011}. 

Semi-analytic models provide an alternative approach to the modeling
of galaxy formation within the framework of a $\Lambda$-cold dark
matter cosmology. Simplified but physically motivated recipes are used
to track the cooling of hot gas into galaxies, the radial size of
discs, the formation of stars, the energy input from supernovae (SNe)
and active galactic nuclei (AGN) into the ISM. The latest generation
of models explicitly includes the detailed tracking of the atomic and
molecular hydrogen content of galaxies and a more physically-motivated
\h2-based star formation recipe
\citep[Somerville, Popping \& Trager in prep.]{Fu2010,Lagos2011cosmic_evol,Lagos2011sflaw,krumholz2011,Fu2012,Popping2013}. These
models have proven to be successful in reproducing the available
observational estimates of the overall \hi and \h2 properties of local
and high-redshift galaxies, such as \hi and \h2 mass, the size-stellar
mass relation of galaxies, \hi and \h2 mass function, the sizes of
\hi discs in the nearby Universe and the observed size evolution of
star-forming discs. Although very useful in shaping our
understanding of galaxy evolution, they do not provide directly
observable predictions of detailed properties of the cold gas in discs
such as temperature and density.  A first attempt to model the CO line emission of galaxies over a wide range of galaxy properties was
made by \citet{Lagos2012}. The authors did not focus on other sub-mm
emission lines.

In this paper we present new predictions of sub-mm line intensities
for galaxies selected from a semi-analytic model of galaxy
formation. We focus on CO line transitions, HCN, [CII] 158 $\mu$m,
[OI] 63 $\mu$m, and the carbon fine structure lines (CI($^3P_1
\rightarrow\,^3P_0$) and CI($^3P_2 \rightarrow\,^3P_1$), hereafter CI
(1-0) and CI (2-1), respectively).  We represent a galaxy by a
three-dimensional distribution of molecular clouds (PDRs) at parsec
scale resolution and then apply radiative transfer and line-tracing
calculations to each molecular cloud to calculate the emitted line radiation of chemical
species by the galaxy. (We do not discuss thermal dust
continuum properties that dominate the broad-band sub-mm
emission of galaxies).
We use these models to explore the sub-mm line-properties of galaxies
at $z =2.0$ and $z = 1.2$, how they shape the star-formation in galaxies, how
these compare to similar galaxies in the local Universe, and to make
qualitative predictions for future ALMA surveys. We aim to understand,
from a theoretical perspective, how the physics that drives the
formation of stars in galaxies during the SF peak of the Universe compares to star
formation in local galaxies and how we can test this. We limit
our study to normal star-forming galaxies, representative of average galaxies on the star-forming ``main sequence'',
rather than focusing on the more extreme starburst
galaxies. Galaxies at $z=2.0$ belong to the
population during the peak of SF activity of the Universe.  Galaxies at $z=1.2$ belong to the tail-end of the
population responsible for the SF that likely produced the bulk of
stars in massive galaxies. Star-forming galaxies at these redshifts
are very favorable for ALMA observations, since one
can observe multiple CO lines, HCN and (both) carbon fine structure lines at
these redshifts with a limited number of bandpasses.  The [CII] and
high-J CO lines can be observed towards galaxies at redshift $z=2.0$.

The structure of the paper is as follows. In Section \ref{sec:model}
we present the theoretical modeling tools used to obtain our
predictions. In Section \ref{sec:results} we present our predictions
for the sub-mm line properties of local and high-redshift galaxies and
how they correlate with typical galaxy properties. In Section
\ref{sec:discussion} we discuss our findings and we summarize our work
in Section \ref{sec:summary}. Throughout this paper we adopt a flat
$\Lambda$CDM cosmology with $\Omega_0 = 0.28,\,\Omega_\Lambda = 0.72,
h = H_0/(100\, \rm{km}\,\rm{s}^{-1}\,\rm{Mpc}^{-1})=0.70,\,\sigma_8 = 0.812$ and a
cosmic baryon fraction of $f_b = 0.1658$ \citep{Komatsu2009}.

\begin{figure*}
\includegraphics[width = \hsize]{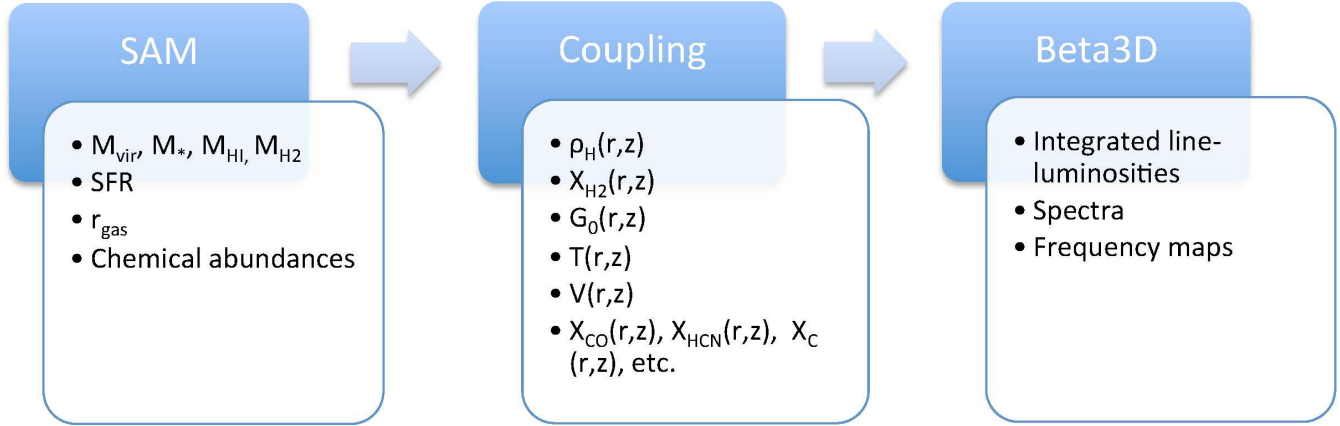}
\caption{Flowchart of our three-step model to calculate the sub-mm
  line intensity from atomic and molecular species in galaxies. The
  blue boxes represent the three individual parts of the model. The
  white boxes show the relevant output from each individual step which
  also acts as input for the following step. \label{fig:flowchart}}
\end{figure*}

\section{Model}
\label{sec:model}
Our model to study the sub-mm properties of galaxies consists of three
parts. The first part of the machinery is a semi-analytic galaxy
formation model (see Section \ref{sec:sam}), which we use to create a
mock sample of galaxies with global properties such as cold gas mass,
\h2 fraction, SFR, and size. The second part is made up of a series of
routines to create 3D realizations of the galaxies in the mock sample
(see Section \ref{sec:coupling}), consisting of key properties
such as density distribution, temperature, and velocity field. The final
part of the machinery is a three-dimensional radiative-transfer and
line tracing code ($\beta$3D; see Section \ref{sec:beta3d}), which
provides integrated luminosities, spectra and frequency maps of
molecular and atomic lines for our mock sample of galaxies. The
overall process, including the output of each step which also acts as
input for the following step, is depicted in Figure
\ref{fig:flowchart}. We provide a summary of the important free
parameters associated with $\beta$3D and the coupling between the
galaxy formation model and $\beta$3D in Table \ref{tab:parameters}. In
the following sections we discuss the individual elements of the
machinery in detail.

\subsection{Galaxy formation model}
\label{sec:sam}
The backbone of our galaxy formation model to create mock galaxies is
the SAM described in \citet{Popping2013}, \citet{Somerville2008} and \citet{Somerville2012}. The model tracks
the hierarchical clustering of dark matter halos, radiative cooling of
gas, star formation, SN feedback, AGN feedback (in two distinct modes,
quasars and radio jets), galaxy mergers, starbursts, the evolution of
stellar populations, and the effects of dust obscuration. We refer the
reader to the above-mentioned papers for a detailed description of the
model and leave the free parameters associated with the
galaxy-formation model fixed at the values discussed in
\citet{Popping2013}. In the remaining of this section we discuss the
physical processes in the model relevant to this work.

When gas cools onto a galaxy, we assume it initially
collapses to form a rotationally-supported disc. The scale radius of
the disc is computed based on the initial angular momentum of the gas
and the halo profile, assuming that angular momentum is conserved and
that the self-gravity of the collapsing baryons causes contraction of
the matter in the inner part of the halo
\citep{Blumenthal1986,Flores1993,Mo1998}.
Assuming that the   halo initially has a density profile described by the
  Navarro-Frenk-White \citep[NFW;][]{Navarro1996} form, the size of
  the gas disc of a galaxy is given by
\begin{equation}
R_{\rm g} = \frac{1}{\sqrt{2}}f_j\lambda R_{\rm vir}f_c^{-1/2}f_{\rm R}(\lambda,c,f_d),
\end{equation}
where $\lambda$ is the halo spin parameter (at redshift zero the
distribution of $\lambda$ typically has a mean of $\sim 0.035$ with a
standard deviation of $0.5$), $f_j \equiv (J_d/m_d)/(J_h/M_{\rm{vir}})$ is the ratio of the
specific angular momentum of the disc and the halo, $c$ is the NFW
concentration of the halo, and $f_d$ is the disc mass to the halo mass
ratio. The parameter $f_c^{-1/2}$ corrects for the difference in energy
of the NFW profile relative to that of a singular isothermal profile,
and $f_{\rm R}$ accounts for the adiabatic contraction \citep[see][for
  expressions governing $f_{\rm R}$ and $f_c$]{Mo1998}. \citet{Somerville2008size}
showed that this approach produces good agreement with the evolution
of the size-stellar mass relation for disc-dominated galaxies from
$z\sim 2$ to the present. This approach also reproduces the observed
sizes of \hi discs in the nearby Universe, the observed sizes of CO
discs in local and high-redshift galaxies, and the spatial extent of the SFR density in
nearby and high-redshift galaxies \citep{Popping2013}.

We assume that the cold gas consists of an ionized, atomic and
molecular component. The \h2 component acts as an important collision
partner and may form in dense regions were the gas is self-shielded
against impinging radiation. The cold gas is distributed in an
exponential disc with radius $r_{\rm gas}$. We divide the cold gas in
radial annuli and compute the fraction of \h2 in each annulus as
described below.

We compute the \h2 fraction of the cold gas in an annulus based on the
simulation by \citet[GK]{Gnedin2011}, who performed high-resolution
``zoom-in'' cosmological simulations with the Adaptive Refinement Tree
(ART) code of \citet{Kravtsov99}, including gravity, hydrodynamics,
non-equilibrium chemistry, and 3D on-the-fly radiative-transfer. Based
on their simulations, the authors find a fitting function for the \h2
fraction which effectively parameterizes $f_{\rm H2}$ as a function of
dust-to-gas ratio relative to the Milky Way, $D_{\rm MW}$, the UV
ionizing background relative to the Milky Way, $U_{\rm MW}$, and the
neutral gas surface density $\Sigma_{HI+H_2}$. The fraction of
molecular hydrogen is given by
\begin{equation}
 f_{H_2} = \left[1+\frac{\tilde{\Sigma}}{\Sigma_{HI+H_2}}\right]^{-2} 
\end{equation}
where
\begin{eqnarray*}
\tilde{\Sigma}  & = &  20\, {\rm M_\odot pc^{-2}} \frac{\Lambda^{4/7}}{D_{\rm MW}} 
\frac{1}{\sqrt{1+U_{\rm MW} D_{\rm MW}^2}}, \\
\Lambda & = & \ln(1+g D_{\rm MW}^{3/7}(U_{\rm MW}/15)^{4/7}),\\
g & = & \frac{1+\alpha s + s^2}{1+s},\\
s &  = & \frac{0.04}{D_*+D_{\rm MW}},\\
\alpha &  = & 5 \frac{U_{\rm MW}/2}{1+(U_{\rm MW}/2)^2},\\
D_* & = & 1.5 \times 10^{-3} \, \ln(1+(3U_{\rm MW})^{1.7}).
\end{eqnarray*}
The local UV background relative to the Milky Way is set by relating
the SFR of the galaxy in the previous time step to the Milky Way SFR
as $U_{\rm MW} = \frac{SFR}{SFR_{\rm MW}}$, where we choose $SFR_{MW} =
  1.0\,M_\odot\,\rm{yr}^{-1}$ \citep{Murray2010,Robitaille2010}. We take the dust-to-gas ratio to be proportional to
the metallicity in solar units $D_{\rm MW} = Z/Z_{\odot}$.

Star formation is modeled based on the empirical relationship found by
\citet{Bigiel2008} between SFR surface density and \h2 surface
density. Observations of high density environments
in starbursts and high-redshift objects indicate that above some
critical \h2 column density, the relationship between SFR surface
density and \h2 surface density steepens. To account for this steepening, we introduce a two-part
process to form stars, 
\begin{equation}
\label{eqn:bigiel2}
\Sigma_{\rm SFR} = A_{\rm SF} \, \Sigma_{\rm H_2}/(10 M_\odot {\rm pc}^{-2}) \left(1+
\frac{\Sigma_{H_2}}{\Sigma_{\rm H_2, crit}}\right)^{N_{\rm SF}},
\end{equation}
with $A_{\rm
  SF}=4.4 \times 10^{-3}\, M_\odot {\rm yr}^{-1} {\rm kpc}^{-2}$,
$\Sigma_{\rm H_2, crit} = 100 M_\odot$ pc$^{-2}$, and $N_{\rm SF}=0.5$.

We use the approach presented in \citet{Arrigoni2010} to include
detailed metal enrichment by type Ia and type II SN and long lived
stars. The chemical enrichment extension allows the tracking of
stellar and gas abundances of individual chemical elements. Although
this extension was originally developed with the
\citet{Somerville2008} models as its backbone, we have incorporated
this machinery into the newest version of the SAMs used in this
paper. We refer the reader to \citet{Arrigoni2010} for a detailed
description of the chemical enrichment model.

\begin{table*}
\caption{Summary of the free parameters associated with the
  three-dimensional radiative-transfer and line tracing code, and its
  coupling with the galaxy formation model.\label{tab:parameters}}
\begin{tabular}{llcl}
\hline
Parameter & Description & Value & Reference\\
\hline
$L_{\rm grid}$ & Grid size & 180 pc & Sec. \ref{sec:densities}\\
$R_{\rm extent}$ & Radius out to which disc is integrated & $4.5
\times R_{\rm g}$ & Sec. \ref{sec:densities}\\
$G_{\rm background}$ & Uniform background radiation field& $1.6\times10^{-3}\,\rm{erg}\,\rm{cm}^{-2}\,\rm{s}^{-1}$ (Habing Flux) &
Sec. \ref{sec:rad_field}\\
$\sigma_{\rm gas}$ & Vertical velocity dispersion of cold gas &
$10\,\rm{km}\,\rm{s}^{-1}$ & \citet{Leroy2008}\\
%$\rho_0$ & Characteristic density of the lognormal PDF $1.0\,\rm{cm}^{-3}$ & \citet{Wada2007}\\
\hline
\hline
\end{tabular}
\end{table*}

\subsection{Coupling the codes}
\label{sec:coupling}
Although SAMs are suitable for modeling the global
properties of galaxies, they lack the spatial information to properly
model the contribution of individual molecular clouds to the emitted
sub-mm spectra of galaxies. In this subsection we describe our
methodology for creating a three-dimensional representation at
parsec-level resolution of model galaxies generated using our SAMs. We
will discuss the galaxy properties that are needed as input for the 3D
radiative-transfer and line tracing code discussed in Section
\ref{sec:beta3d}.

\subsubsection{Gas densities}
\label{sec:densities}
For an exponential disc, the effective central hydrogen gas density seen from any radius $r$ in the disk is defined as
\begin{equation}
  n_0(r) = \frac{M_H}{4\pi m_H\,R^2_g z_g(r)}
\end{equation}
where $M_H$ is the total hydrogen mass (atomic plus molecular) of the
galaxy, $m_H$ the mass of a single
hydrogen atom, $R_g$ the gas scale length of the galaxy and $z_g(r)$ the
gas scale height. 

The scale height $z_g(r)$ of the gaseous disc is calculated assuming
vertical equilibrium, where the gravitational force is balanced by
the pressure of the gas $P(r)=\sigma_{\rm{gas}}^2\,\rho_{\rm g}(r)$,
where $\rho_{\rm g} = \Sigma_{\rm gas}(r)/2z_g(r)$, and $\sigma_{\rm{gas}}$
is the vertical velocity dispersion of the gas in the disc which we
assume to be constant, $\sigma_{\rm{gas}} = 10\,\rm{km}\,\rm{s}^{-1}$
\citep{Leroy2008}. Following
\citet{Popping2012} we can express the pressure acting on the disc
as 
\begin{equation}
P_m(r) =
\frac{\pi}{2}\,G\,\Sigma_{\mathrm{gas}}(r)\bigl[\Sigma_{\mathrm{gas}}(r) + \frac{\bar f_\sigma}{4}\sqrt{\Sigma_*(r)\Sigma_*^0}\bigr],
\end{equation}
where $\bar f_\sigma = 0.4$, $\Sigma_*(r)$ the stellar surface density and
$\Sigma_*^0$ the central stellar surface density. We can now solve for
$z_g$ to find
\begin{equation}
z_g(r) = \frac{\sigma_{\rm{gas}}^2}{\pi\,G\,\bigl[\Sigma_{\mathrm{gas}}(r) + 0.1\sqrt{\Sigma_*(r)\Sigma_*^0}\bigr]}.
\end{equation}
The hydrogen density at any point in the galaxy can now be expressed as
\begin{equation}
\label{eq:exponential}
n_H(r,z) = n_0(r)\,\exp\left(-\frac{r}{R_g}\right)\exp\left(-\frac{|z|}{z_g(r)}\right).
\end{equation}
This approach is similar to that presented in \citet{Berry2013},
although we do not assume the ratio between gas scale height and scale
length $\chi_z$ to be fixed, but rather calculated the gas scale
height at every point in the disc. We adopt a resolution of 180 pc and integrate the discs
out to 4.5 times their scale radii.

\subsubsection{\h2 abundance}
\label{subsec:h2_abund}
The local \h2 abundance of cold gas is dependent on the local cold gas
(column) density, whereas SAMs only provide the global \h2 abundance.
We therefore need to recalculate the \h2 abundance in every
grid cell. The self-shielding transition of \h2 is rather sharp for an
impinging UV-radiation field in a PDR \citep[see equations 36 and 37
  of][]{Draine1996}. We therefore may assume that the cold gas in
grid cells with a visual extinction of $A_v>2.0$ mag is fully
molecular (i.e., the fractional abundance $X_{H2}=0.5$) and for $A_v <
1.0$ mag one has the minimal \h2 abundance as suggested by the
SAMs. The visual extinction of a grid cell is defined as $A_v = N_{\rm
  H}/2\times10^{21} (\mathrm{mag})$, where $N_{\rm H}$ is the hydrogen
column density of the grid cell. We ascribe \h2 abundances to
grid cells with $1.0 < A_v < 2.0$ mag by logarithmic interpolation
between the abundances set for visual extinction magnitudes $A_v =
1.0$ mag and $A_v = 2.0$ mag.

\subsubsection{Impinging radiation field}
\label{sec:rad_field}
The radiation field $G_{UV}$ impinging on a giant molecular cloud (GMC) depends on the transmission of UV
photons from star-forming regions and their propagation through the
diffuse ISM. The exact UV field strength depends on the local conditions in
the ISM, such as optical depth and the ratio of gas and dust in the diffuse ISM
and in GMCs. We scale $G_{UV}$ as a function of the local
star-formation rate in each individual grid cell. The SFR
within a grid cell is calculated following Equation \ref{eqn:bigiel2}.
We can convert the local SFR to a UV radiation field $G_{UV}$
by relating the SFR surface density within a grid cell to the UV
radiation-field as
\begin{equation}
\frac{G_{UV}}{G_0} =
\left(\frac{\Sigma_{\rm{SFR}}}{\Sigma^0_{\rm{SFR}}}\right).
\end{equation}
We choose $\Sigma^0_{\rm{SFR}} =
10^{-3}M_\odot\,\rm{yr}^{-1}\,\rm{kpc}^{-2}$, to assure $G_{UV} =
G_0 = 1.6\times10^{-3}\,\rm{erg}\,\rm{cm}^{-2}\,\rm{s}^{-1}$ (the
Habing Flux) for the solar neighborhood. We also include a uniform local background
radiation field of 1 $G_0$. 

\subsubsection{Temperature}
The temperature within a PDR is set by the balance between heating and
cooling processes. Cooling of PDRs occurs predominantly through
fine-structure lines of abundant atoms and ions. In the moderately dense ISM  ($n < n_{\rm{cr}} \approx 1\times 10^4\, \rm{cm}^{-3}$), the cooling is
mainly dominated by [CII], whereas the [OI] 63 $\mu m$ fine-structure
line is important in higher density regions. \citet{Tielensbook} shows
that in the high density limit (where [OI] is the dominant coolant)
the energy balance of a steady system can be written as
\begin{equation}
\frac{\exp(-y)}{y^{1.17}} = \frac{3.4 \gamma}{1 + 2.5\gamma^{0.73}},
\end{equation}
where
\begin{equation}
y = \frac{228 \,K}{T}
\end{equation}
and $\gamma$ is the ionization parameter defined as $G_0T^{1/2}/n$ with
$n_e= 1.4\times 10^{-4}n$.  $\gamma$ measures the ionization rate over
the recombination rate, the efficiency of the photo-electric
effect. These expressions can be solved iteratively for the
temperature $T$ for a given density $n$ and impinging radiation field $G_{UV}$.

In the lower density regime, where [CII] is the dominant coolant, we
can solve for the temperature using a similar set of equations,
\begin{equation}
\frac{\exp(-y')}{y'^{1/2}} = \frac{1.7 \gamma}{1 + 2.5\gamma^{0.73}}
\end{equation}
where
\begin{equation}
y' = \frac{92 \,K}{T}.
\end{equation}

Up to this point we have neglected the contribution to the total
cooling through molecules (especially CO). CO acts as a good coolant
for cold gas due to its ability to radiate at relatively low
temperatures and densities. However, the cooling rate is difficult to
calculate because lower-J CO lines are optically thick. A photon
emitted by a molecule in the J=1 state is likely to be re-absorbed by
another molecule in the J=0 state. This process effectively only moves
the energy around within the molecular cloud without contributing to
the overall net cooling. Molecules in these low states only contribute
to the cooling when they are within approximately one optical depth
from the cloud surface, restricting the cooling only to a small
fraction of the cloud volume. Emission from high-J levels of CO is
optically thin and can escape the cloud, however, the temperatures and
densities required to excite these molecules up to high-J levels are
generally high ($\sim\,10^5\,\rm{cm}^{-3}$ and $\sim 100$ K). 
In the environments typical of normal star forming galaxies, there are
therefore only a few molecules in these high states, which strongly
suppresses cooling from these molecules.

\subsubsection{Velocity Field}
In addition to the density and temperature of the cold gas, the 3D
radiative-transfer simulation requires the velocity field of the gas as one
of its initial conditions for the line-tracing. Furthermore it allows the creation of
frequency or velocity maps of sub-mm emission when a velocity field of
the disc is provided \citep{JP2011}. We derive the velocity field
following the approach presented in \citet{Obreschkow2009_sam}. We
refer the reader to that work for a detailed description of the
methodology, but present the main ingredients of their approach.  The
circular velocity profile of gas in a disc comprises three components, the halo, the galaxy disc and the bulge of the galaxy,
\begin{equation}
V_c^2(x) = V^2_{c, {\rm{halo}}}(x) + V^2_{c, \rm{disc}}(x) + V^2_{c, \rm{bulge}}(x),
\end{equation}
where $x = r/r_{\rm{vir}}$, and $r_{\rm{vir}}$ is the virial radius of
the halo. Each galaxy is situated in a dark matter halo with an NFW profile. The concentration parameter of the halo
$c_{\rm{halo}}$, as well as its virial radius $r_{\rm vir}$, are both
provided by the galaxy formation model. Assuming a spherical halo, the circular
velocity of the gas contributed by the halo mass is given by
\begin{equation}
V^2_{c,\rm{halo}}(x) = \frac{G\,M_{\rm{vir}}}{r_{\rm vir}}\times \frac{\ln{(1+ c_{\rm{halo}}x)}  - \frac{c_{\rm{halo}}x}{1 + c_{\rm{halo}}x}}{x[\ln{(1+ c_{\rm{halo}}x)}  - \frac{c_{\rm{halo}}}{1 + c_{\rm{halo}}}]},
\end{equation}
where $M_{\rm{vir}}$ is the virial mass of the halo.

Under the assumption that the radial density distribution of a galaxy
disc can be described by an exponential,
\citet{Obreschkow2009_sam} find that the circular velocity of cold gas
due to the disc can be approximated by
\begin{multline}
V^2_{c, \rm{disc}}(x) \approx
\frac{GM_{\rm{disc}}}{r_{\rm{vir}}}\times \\ \frac{c_{\rm{disc}} + 4.8 c_{\rm{disc}}\exp[-0.35 c_{\rm{disc}}x -
  3.5/(c_{\rm{disc}}x)]}{c_{\rm{disc}}x + (c_{\rm{disc}}x)^{-2} +
  2(c_{\rm{disc}})^{-1/2}},
\end{multline}
where $c_{\rm{disc}} = r_{\rm{vir}}/r_{\rm{disc}}$ and $M_{\rm{disc}}$
the sum of the stellar and gaseous component of the disc.

Following \citet{Obreschkow2009_sam}, we assume that the
bulges of all galaxies are spherical, and that their density profiles
can be described by a Plummer potential \citep{Plummer1911}, with a characteristic radius $r_{\rm{Plummer}}\approx
1.7\,r_{\rm{bulge}}$ and $r_{\rm{bulge}} \approx 0.05\,r_{\rm{disc}}$. The circular velocity profile component due to
the bulge is
\begin{equation}
V^2_{c, \rm{bulge}}(x) = \frac{GM_{\rm{bulge}}}{r_{\rm{vir}}}\times
  \frac{(c_{\rm{bulge}}x)^2 c_{\rm{bulge}}}{[1 + (c_{\rm{bulge}}x)^2]^{3/2}},
\end{equation}
where $c_{\rm{bulge}} = r_{\rm{vir}}/r_{\rm{Plummer}}$ and
$M_{\rm{bulge}}$ is the mass of the bulge.

With an expression for the circular velocity profiles of the halo, disc,
and bulge, we can now calculate the circular velocity of the cold gas
at any location in the disc. The vertical velocity dispersion of the
cold gas $\sigma_{\rm{gas}}$ is responsible for the motions of the gas perpendicular to the
disc. Observations in the local Universe have shown that the vertical
velocity dispersion of the cold gas $\sigma_{\rm gas} \approx 10
\,\rm{km}\,\rm{s}^{-1}$ and is approximately constant over the disc \citep[e.g.][]{shostak1984,Leroy2008}. We ascribe a vertical velocity to each
grid cell by randomly picking a velocity from a Gaussian distribution
centered around $0\,\rm{km}\,\rm{s}^{-1}$ with a standard deviation of
$10\,\rm{km}\,\rm{s}^{-1}$. We estimate the local turbulent velocity
dispersion within a GMC following \citet{Larson1981}, evaluated
for the physical size of a grid cell in the model.

\subsubsection{Abundances}
\label{sec:abundances}
Good estimates of the abundances of the various species are
fundamental to properly model the line contribution from atomic and
molecular species. We already discussed the \h2 abundance of the cold
gas in detail in Section \ref{subsec:h2_abund}; however, we also need
to take into account the atoms, molecules and ions of interest to us.

The latest version of our SAMs includes detailed tracking of the
abundance of single elements in the cold gas \citep{Arrigoni2010}. We use this model to
calculate the carbon and oxygen abundance of the ISM.

The CO abundance of the cold gas is calculated as the amount of carbon locked up in
CO.  The fraction of the carbon mass locked up in
CO has an explicit dependence on metallicity. Following  \citet{Wolfire2010} we
calculate this fraction as 
\begin{multline}
f_{\rm CO} = f_{\rm H2} \times \\ 
\exp{[-4(0.52 - 0.045\ln{\frac{G_{UV}/G_0}{n_H}} - 0.097\ln{Z'})/A_v]},
\end{multline}
where $Z'$ is the metallicity of the cold gas expressed in solar units. 

We adopt an HCN abundance of $10^{-8}$ with
respect to the molecular gas, and scale the C$^+$ abundance with the
abundance of carbon in the cold gas. These choices yield good agreement with predictions from
\citet{Meijerink2005} for the typical range of densities and radiation
fields relevant to our work.

\subsection{3D radiative-transfer and line tracing code}
\label{sec:beta3d}
We use the advanced fully three-dimensional radiative-transfer code
$\beta$3D \citep{Poelman2005,Poelman2006} optimized for heavy memory
usage by \citet{JP2011}. The optimized version was initially developed
to calculate the three-dimensional transfer of line radiation in 256 x
256 x 128 element data cubes at a spatial resolution of 0.25 pc, but
works as well for different shaped data cubes at different
resolutions.

The code requires density, \h2 abundance, temperature, impinging
radiation field, and velocity as ambient conditions in each grid cell
to calculate the transfer of line radiation of molecules by use of a
non-local escape probability formalism.  We refer the reader to
\citet{Poelman2005} for a detailed description of the
radiative-transfer method.

Level populations of different atomic and molecular species (in this
study C$^+$,C,O, $^{12}$CO and HCN) are calculated using collision
rates available in the LAMDA database
\citep{Schoier2005}. We use \h2 as the main collision partner for the
radiative transfer calculations of all the molecules and also include
the contribution of helium atoms to the total collision density for
CO and HCN \citep[following][]{JP2011}. We include the contribution
from atomic hydrogen and electrons for the collision rates of C, O,
and C$^+$. The densities of the collisional partners are derived from
the galaxy formation model.

\begin{figure*}
  \includegraphics[width = 0.95\hsize]{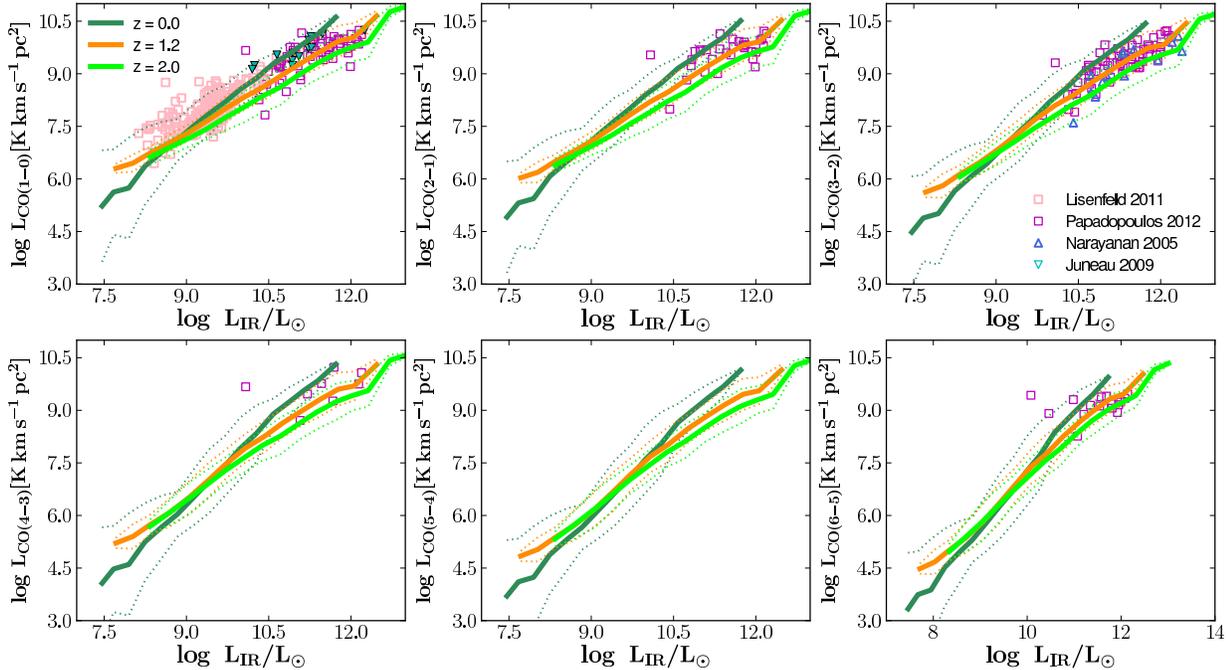}
  \caption{CO line-luminosity of CO $J=1-0$ up to CO $J=6-5$ as a
    function of FIR luminosity. Model results are compared to
    observations taken from \citet{Narayanan2005}, \citet{Juneau2009}, \citet{Lisenfeld2011}, and \citet{Papadopoulos2012}. The solid
    lines show the median of the model predictions, whereas the dotted
    lines represent the two sigma deviation from the median. 
\label{fig:CO_scaling}}
\end{figure*}

The intensity from molecules is dependent on the distribution of level
populations, which depends on the temperature and the density of the
gas (through collisions with other molecules and atoms as well as a
background impinging radiation field). There are large differences in
the collisional densities necessary to excite different energy
levels. For example, the density needed to populate the $J=1$ state of
CO through collisional excitation is $\sim 10^2-10^3\,\rm{cm}^{-3}$,
whereas it is already $\sim 10^4\,\rm{cm}^{-3}$ for the $J=3$
state. These densities arise in different regions of the GMCs, the
former being the diffuse GMC atmospheres and the latter the dense GMC
core. Densities used in the radiative-transfer calculations in this
work were calculated by post-processing the results from the galaxy
formation model, smoothed to a resolution of 180 pc (see Sections
\ref{sec:sam} and \ref{sec:coupling}). At this resolution, the
smoothed density profile of our galaxy never exceeds densities of a
few times $10^3 \rm{cm}^{-3}$, typical for the diffuse outer parts of
the GMC, and a sub-grid approach is necessary to account for
  radiation from denser environments. 

Numerical simulations often describe the
structure of the ISM with a lognormal (LN) probability distribution function (PDF)
\citep[e.g.,][]{Wada1999,Krumholz2005,Wada2007,
  Hennebelle2008, Hennebelle2009,Price2011}:
\begin{equation}
  \label{eq:LNPDF}
  f(\rho)d\rho = \frac{1}{\sqrt{2\pi}\sigma}\exp\left[-\frac{\ln{(\rho/\rho_0)^2}}{2\sigma^2}\right]d\ln{\rho},
\end{equation}
where $\rho_0$ is the characteristic density and $\sigma$ the
dispersion of the distribution.  Numerical simulations suggest that
$\sigma$ can be expressed in terms of the one-dimensional Mach number
of turbulence $M$ \citep[e.g.,][]{Ostriker2001,Lemaster2008,Price2011}:
\begin{equation}
\sigma^2 \approx \ln{(1 + 3M^2/4)}.
\end{equation}
The Mach number is calculated as the ratio between the local turbulent
velocity ($\Delta v_d$) and the sound speed ($c_{\rm s}$) of the medium. The local
turbulent velocity is estimated following \citet{Larson1981}, evaluated
for the physical size of a grid cell in the model. The
volume-averaged density of a LN distribution is given by
\begin{equation}
\langle\rho\rangle_V = \rho_0e^{2\sigma^2}.
\end{equation}
In our case the volume-averaged density of each grid-cell is also given by
Equation \ref{eq:exponential}. We can therefore express the
characteristic density $\rho_0$ as
\begin{equation}
\rho_0 = \frac{\langle\rho\rangle_V}{e^{2\sigma^2}}.
\end{equation}

We assume that a grid-cell is made up by small molecular clouds drawn from the
LN-PDF.  We calculate the contribution of each of the individual molecular clouds
within the sub-grid to the emitted radiation, taking the overlap in
optical depth space of the molecular clouds into account.

Line intensities are computed using a ray-tracing approach, including
the effects of kinematic structures in the gas and optical depth
effects. The emerging specific intensity is computed using the escape
probability formalism presented in \citet{Poelman2005},
\begin{equation}
dI^z_\nu = \frac{1}{4\pi}n_iA_{ij}h\nu_{ij}\beta(\tau_{ij})\,\bigl(\frac{S_{ij}-I_b^{loc}(\nu_{ij})}{S_{ij}}\bigr)\phi(\nu)dz,
\end{equation}
where $dI^z_\nu$ has units of erg cm$^{-2}$ s$^{-1}$ sr$^{-1}$
Hz$^{-1}$, $n_i$ is the population density in the $i$th level,
$A_{ij}$ the Einstein A coefficient, $h\nu_{ij}$ is the energy
difference between the levels $i$ and $j$, $\beta$ the escape
probability of a photon, $\tau_{ij}$ the cumulative optical depth, and
$\phi(\nu)$ the Doppler correction to the photon frequency due to
local turbulence inside the cloud and large scale bulk
motions. $S_{ij}$ is the source function of the corresponding medium,
and $I_b^{loc}(\nu_{ij})$ the local continuum background radiation at
the field frequency $\nu_{ij}$. The local background
$I_b^{loc}(\nu_{ij})$ is caused by the local dust emission
\citep[calculated following][]{Hollenbach1991} and is slowly varying
with frequency.\footnote{We note that this is an extension to the
  model presented in \citet{JP2011}.} The code uses a multi-zone
radiative-transfer approach in which the emerging specific intensity
is dependent on the different escape probabilities within a grid-cell
as well as connecting adjacent grid points along the line of sight. This makes our approach more physical compared to the purely local nature of the LVG approximation.

% The gas within each molecular cloud is treated as a static medium. In
% reality the free-fall time ($t_{\rm{ff}} =
% [3\pi/(32G\rho)]^{\frac{1}{2}}$) may be less than the sound-crossing time
% ($t_{\rm{cross}} = R/c_{\rm{s}}$) and rather than being in a stable
% equilibrium, the system undergoes gravitational collapse with a
% velocity gradient $dV/L \sim 1/t_{\rm{ff}}$. The effective
% contribution from each molecular cloud to the observational signal therefore needs to be
% corrected by the ratio between the free-fall time and the
% sound-crossing time.

\section{Results}
\label{sec:results}
In this section we present the predictions of our model for different
atomic and molecular species. The simulations were run on a grid of
halos with virial masses ranging from $1\times10^9\,M_\odot$ to
$5\times10^{14}\,M_\odot$, with a mass resolution of $1\times 10^7 M_\odot$. Throughout the rest of the paper, we only
consider central galaxies with a bulge-to-total mass ratio of 0.4 and smaller, and gas fractions
of $f_{\rm gas} > 0.1$. We first discuss the predicted CO
line emission from model galaxies and how line ratios can potentially
be used to constrain the gas physics in play in galaxies. We will
continue with similar discussions for HCN, atomic and ionized
carbon, and oxygen, after which we will discuss the cooling properties
of the modeled galaxies.

\begin{figure*}
 \includegraphics[width = 0.95\hsize]{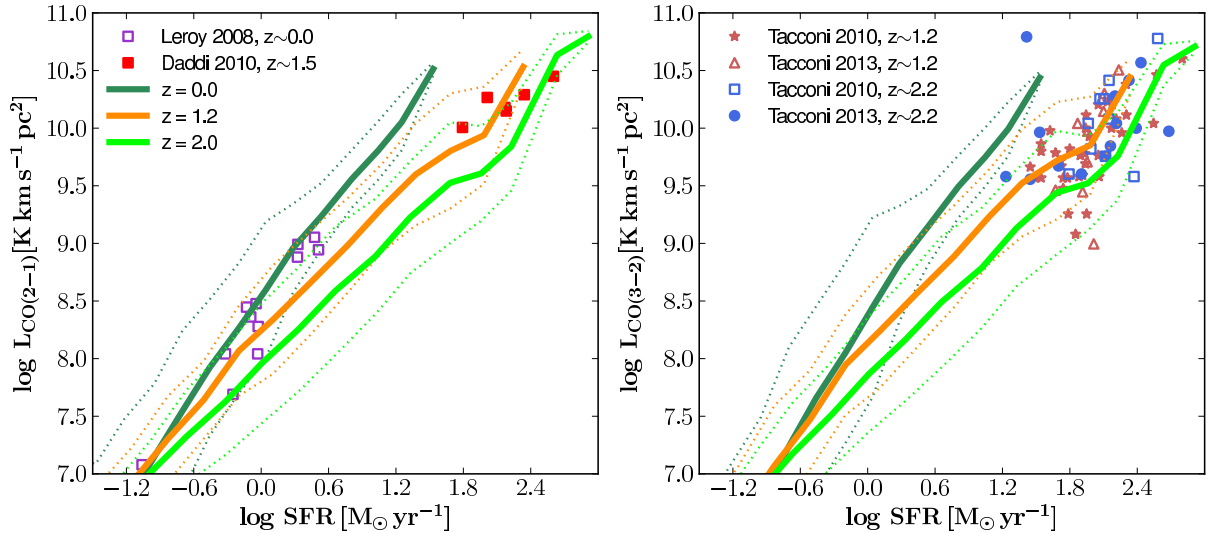}
 \caption{CO line luminosity of CO $J=2-1$ (left panel) and CO $J=3-2$
   (right panel) as a function of SFR for modeled galaxies at $z=0.0$,
   $z=1.2$, and $z=2.0$. Observations are taken from \citet{Leroy2008},
   \citet{Daddi2010}, \citet{Tacconi2010}, and \citet{Tacconi2013}. \label{fig:CO_scaling_SFR}}
 \end{figure*}

\begin{figure*}
  \includegraphics[width = \hsize]{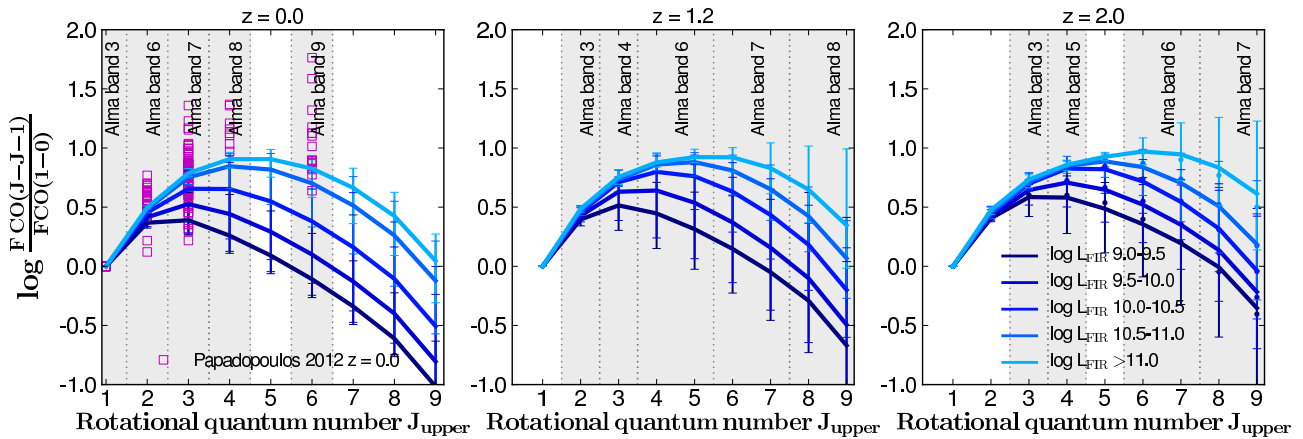}
  \caption{CO spectral line energy distribution (SLED) of our modeled
    galaxies at $z=0.0$ (left panel), $z=1.2$ (middle panel), and
    $z=2.0$ (right panel) separated into bins of FIR luminosity. The
    SLED is normalized to the CO $J=1-0$ line luminosity. Observations
    are taken from \citet{Papadopoulos2012}. Note the change in CO
    SLED shape between galaxies at $z=0.0$, $z=1.2$, and $z=2.0$ \label{fig:CO_SLED}}
\end{figure*}

\begin{figure*}
\includegraphics[width = \hsize]{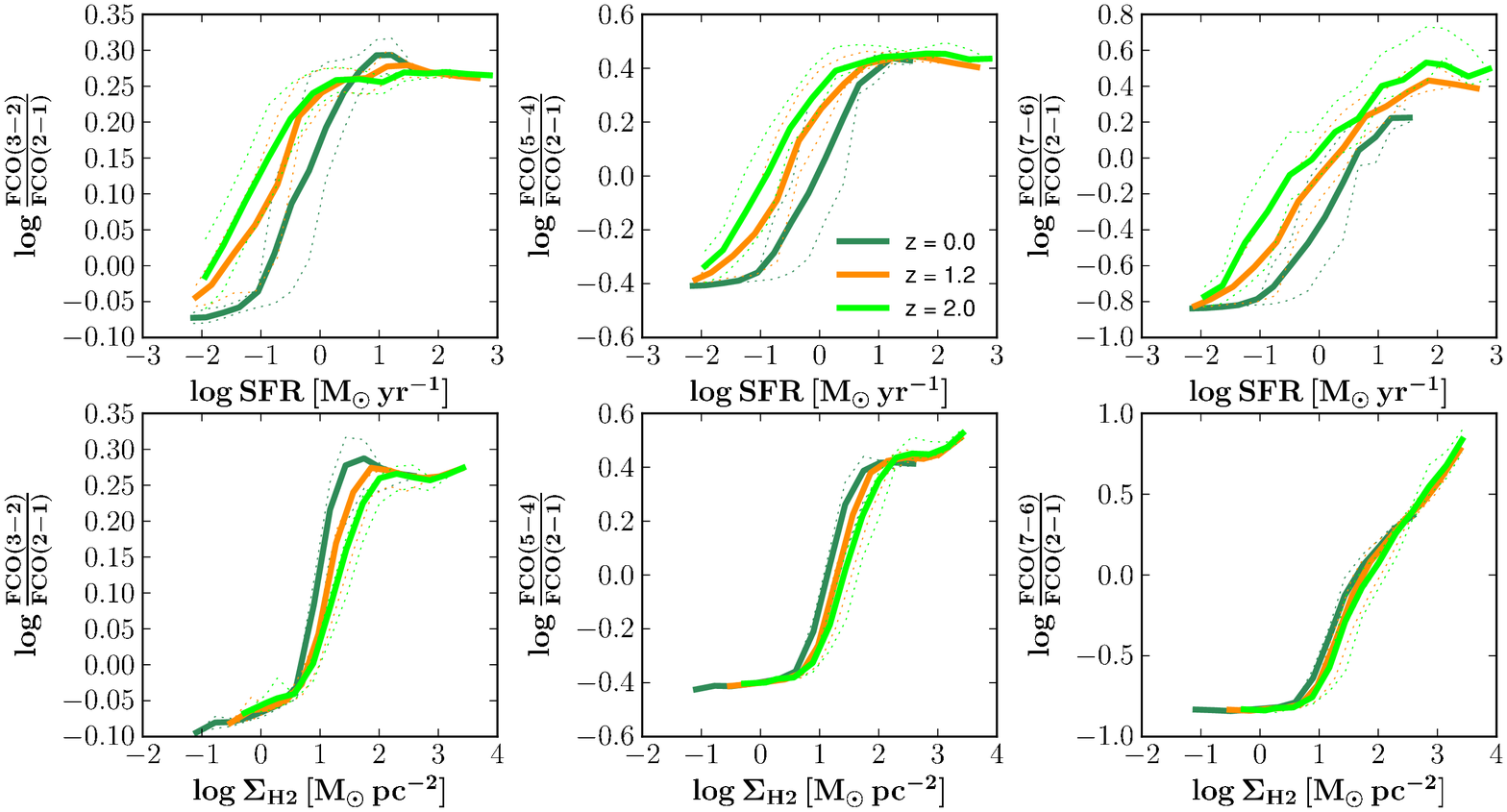}
\caption{CO line ratios as a function of global galaxy properties at
  $z=0.0$, $z=1.2$, and $z=2.0$. CO line ratios of interest are CO $J=3-2$/CO
  $J=2-1$ (left column), CO $J=5-4$/CO $J=2-1$ (middle column), and CO
  $J=7-6$/CO $J=2-1$ (right column). CO line ratios are plotted as a
  function of SFR (top row) and cold gas surface density ($\Sigma_{\rm{HI} + \rm{H2}}$;bottom
  row). Note the offset in line ratios between galaxies at $z=0.0$ and
  higher redshift for the SFR, but the tight correlation in line ratios as a
  function of cold gas surface-density. \label{fig:CO_ratio}}
\end{figure*}

\subsection{CO}
\label{sec:CO}
Because of its relatively high abundance ($\sim\,10^{-4}$), the
emission of CO is commonly used as a tracer of the ISM. Indeed, in the
last decade surveys have probed the CO emission of local galaxies with
a broad range of properties
\citep[e.g.,][]{Helfer2003,Narayanan2005,Leroy2008,Lisenfeld2011,Papadopoulos2012,
  Bauermeister2013}, adding a great wealth of information about the
molecular content of local galaxies. In Figure \ref{fig:CO_scaling} we
present our predictions for CO line luminosities up to CO J$=6-5$ as a
function of FIR luminosity in local galaxies. Observations were taken from
\citet{Narayanan2005}, \citet{Juneau2009}, \citet{Lisenfeld2011}, and
\citet{Papadopoulos2012}. We find that the CO luminosity of galaxies
increases monotonically with increasing FIR luminosity for all CO
transitions. We find a good match between model predictions and
observations over a wide range of FIR luminosities for the CO J$=1-0$
transition, as well as for the higher CO transitions. This is
encouraging, as it implies that we do not only correctly reproduce the
CO $J=1-0$ emission arising from low temperature and low density
regions (a few times $10^3\,\rm{cm}^{-3}$), but also the CO $J=6-5$
emission from much warmer and denser regions (up to $10^5\,\rm{cm}^{-3}$). The CO line luminosity of galaxies at
  fixed FIR luminosity decreases with increasing redshift. We find a shallower slope in the relation between CO line luminosity
and FIR luminosity for galaxies at $z=1.2$ and $z=2.0$ than for local
galaxies.

Figure \ref{fig:CO_scaling_SFR} is similar to the previous figure, but
shows the emitted CO luminosity as a function of SFR. Observations
were taken from \citet{Leroy2008}, \citet{Daddi2010},
\citet{Tacconi2010}, and \citet{Tacconi2013}. Similar to the previous
figure, the CO luminosity increases monotonically with increasing
SFR. We find good agreement with direct observations, both in the
local Universe and at high redshift. Again, we note that the
critical density and excitation energy of CO $J=3-2$ and CO $J=2-1$
differ by approximately a factor of two, demonstrating that our model
correctly describes the gas physics in play over a range of densities
and temperatures.

It is clear from the previous figure (and many of the subsequent
figures) that at fixed FIR luminosity modeled galaxies at $z=1.2$ and $z=2.0$
emit somewhat less radiation through lines tracing diffuse and
low-density environments. We discuss this in Section
\ref{sec:discussion}. The sparse amount of data for typical
SF galaxies at high redshift unfortunately does not allow us to
further constrain our model predictions, both due to a lack of
galaxies with lower SFR and a lack of observations covering a wide
enough range of CO lines. We anticipate that in the near future ALMA
and other sub-mm telescopes will add significantly to the number of high-redshift galaxies with
(multiple) observed CO line-transitions.

The CO Spectral Line Energy Distribution (SLED) of galaxies provides
straightforward information about the density and temperature of the
dominant SF cold gas. Different densities and temperatures result in a
change of peak location in the CO SLED: high CO excitation is achieved
through a combination of high kinetic temperature and high density. In
Figure \ref{fig:CO_SLED} we present the CO SLEDs for our modeled
galaxies, compared to observations of CO line ratios by
\citet{Papadopoulos2012}. The CO SLED is normalized to the CO $J=1-0$
emission. On average our predictions at $z=0.0$ are in good agreement with the
observations. We do not reproduce the highest excitation CO SLEDs from
the observational sample. Papadopoulos et al. argue that supersonic
turbulence and high cosmic ray energy densities are necessary to power
the extraordinary CO line excitation of these galaxies. We have not
included these physical processes in our model. We find that the
more FIR-luminous objects peak at higher CO excitations, corresponding
to higher kinetic temperatures and densities. This implies that the
physical properties of the ISM in the more FIR-luminous objects differ
significantly from the ISM in less FIR-luminous objects and that the
ISM is not just made up by a larger number of clouds with similar
density and temperature.

We see the same trends at high redshifts, where again the FIR-luminous
objects have a higher CO-excitation peak. Furthermore, there is a
strong difference in shape when comparing the $z=0.0$, $z=1.2$, and $z=2.0$ CO
SLEDs at fixed FIR luminosity. For example, galaxies with FIR
luminosities $9.5 < \log{(L_{\rm FIR}/\rm{L}_\odot)} < 10.0$ show a peak in
their CO SLED at the third rotational level, whereas the CO SLED peaks
at the fourth rotational level for similar galaxies at
$z=1.2$ and $z=2.0$, respectively. Analogously, galaxies with FIR
luminosities $10.5 <\log{(L_{\rm FIR}/\rm{L}_\odot)} < 11.0$ peak at the fourth level in
$z=0.0$ galaxies, whereas they peak at the fifth level in
$z=1.2$ and $z=2.0$ galaxies, respectively. There is an even stronger
evolution in the CO SLED for the most FIR-luminous galaxies $\log{(L_{\rm
      FIR}/\rm{L}_\odot)} > 11$. The increment in the excitation
peak with redshift is indicative of
higher densities and kinetic temperatures. Furthermore, the relative
contribution of high excitation CO lines is much larger in the
high-redshift modeled galaxies. This strongly favors a scenario in
which the ISM in galaxies with fixed FIR luminosities at high redshift is
denser and warmer than in their local counterparts. Galaxies at
$z=2.0$ consist of denser and warmer gas than their counterparts at $z=1.2$.

For a fixed density and temperature a change in CO abundance
can affect the shape of the CO SLED. The CO SLED of a molecular
cloud peaks at higher rotational levels with increasing CO
abundance. To first order the CO abundance of the molecular gas in our
galaxies follows the cold gas metallicity, which at fixed FIR
luminosity increases with time. If the excitation
conditions (gas density and temperature) of galaxies at high
redshift would be similar to local galaxies, we would expect the peak
in the CO SLED of high-redshift galaxies to move to lower rotational
levels due to lower CO abundances. This is in sharp contrast with our predictions,
emphasizing that the change in CO SLED with redshift is driven by different
excitation conditions.

In Figure \ref{fig:CO_ratio} we explore the line ratios of CO
transitions as a function of two global galaxy properties. We compare our
predicted line ratios with SFR and \h2 surface-density. We find a strong
increase in CO line ratios as a function of SFR. Most notable is that
the increase is more significant when using line ratios tracing larger
differences in density and temperature. Whereas we only observe an
increase of approximately 0.5 dex in the CO($J=3-2$)/CO($J=2-1)$ line
ratio, we find an increase of a full dex in the
CO($J=5-4$)/CO($J=2-1)$ ratio, and two dex in the
CO($J=7-6$)/CO($J=2-1)$ ratio. This indicates that with increasing
SFR the cold gas contains a much larger fraction of gas clouds with
densities of the order $\sim 10^5\,\rm{cm}^{-3}$. Galaxies at $z=1.2$
and $z=2.0$ have higher CO line ratios than their counterparts with similar SFR
at $z=0$. This is most prominent for galaxies with $\rm{SFR} < 10
\rm{M}_\odot\,\rm{yr}^{-1}$.The difference between the local and
high-redshift galaxies also increases when using line ratios tracing
larger differences in density and temperature.

The evolution in CO line ratios with \h2 surface density is similar to that with SFR: line ratios increase with increasing FIR
luminosity, and the differences become more prominent for line ratios
tracing higher densities. The offset in line ratios between high
redshift and local galaxies, on the other hand, is negligible. This suggests
that CO line ratios are indeed reliable tracers of the density of the
dominant gas population of a galaxy, independent of redshift.

\begin{figure}
\includegraphics[width = \hsize]{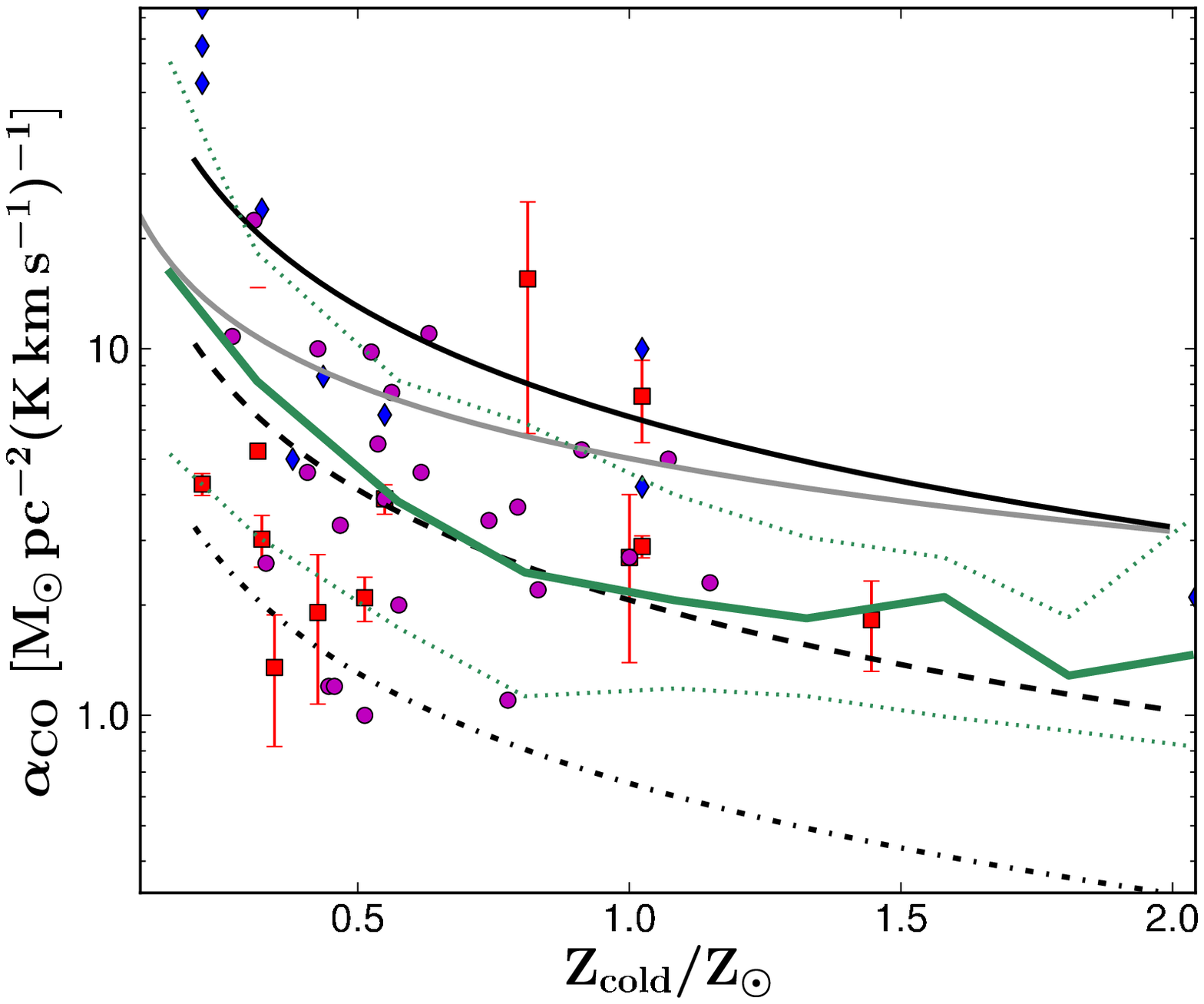}
\caption{The CO-to-\h2 conversion factor $\alpha_{\rm CO}$ in galaxies
    at $z=0.0$ as a function of cold gas metallicity (green solid
    line). The dotted lines mark the $2\sigma$ deviation from the
    mean. Observational measures of $\alpha_{\rm CO}$ were taken
    from \citet[red squares]{Bolatto2008}, \citet[blue
    diamonds]{Leroy2011}, and \citet[purple
    circles]{Sandstrom2012}. Theoretical predictions were taken from
    \citet[grey solid line]{Feldmann2012} and \citet{Narayanan2012}
    for an \h2 surface density of 10 (black solid), 100 (black
    dashed) and 1000 (black dash-dotted) $M_\odot\,\rm{pc}^{-2}$.\label{fig:alpha_CO}}
\end{figure}

\begin{figure*}
 \includegraphics[width = \hsize]{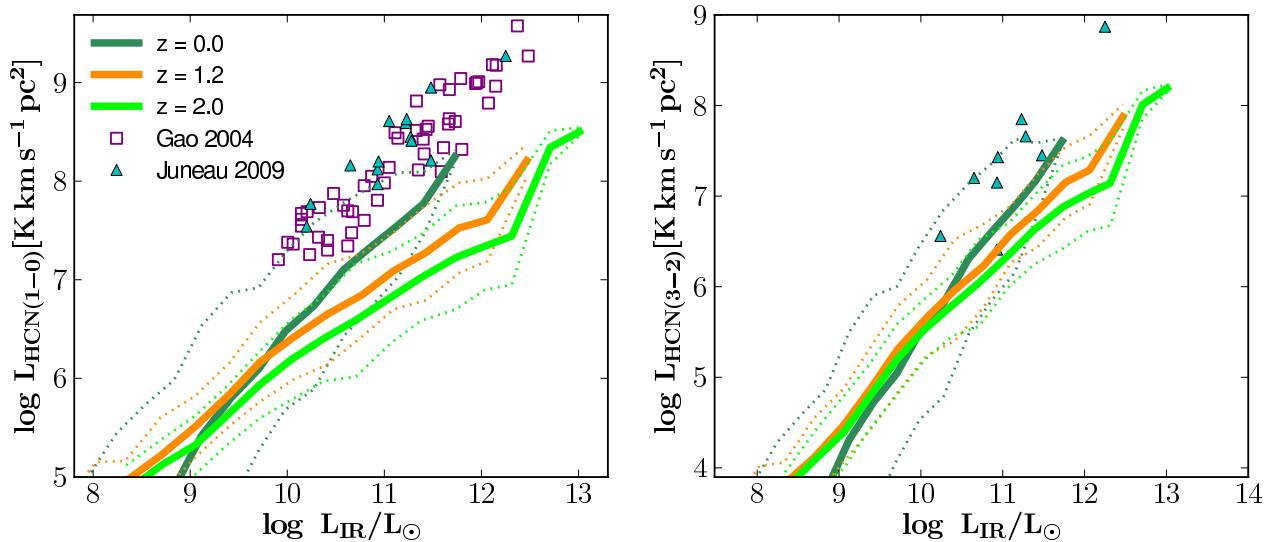}
\caption{The HCN $J=1-0$ (left panel) and HCN $J=3-2$ (right panel)
  luminosity of galaxies as a function of its FIR luminosity for
  galaxies at $z=0.0$, $z=1.2$, and $z=2.0$. Observations of local galaxies are
  taken from \citet{Gao2004} and
  \citet{Juneau2009}.\label{fig:scale_HCN}}
\end{figure*}

\subsubsection{CO-to-\h2 conversion}
The most common way to estimate the molecular gas mass of a galaxy is
through the conversion factor between CO $J=1-0$ luminosity and \h2
mass,
\begin{equation}
\alpha_{\rm CO} = \frac{M_{\rm H2}}{L_{\rm CO}}.
\end{equation}
Although the CO-to-\h2 conversion appears to be constant in our own
Milky Way, there is substantial evidence that this conversion departs
from the standard Milky Way value in low-metallicity and high surface
density environments \citep[e.g.,][]{Schruba2011,Genzel2012}. Indeed,
several efforts have been carried out to model $\alpha_{\rm CO}$ as a
function of the metallicity and \h2 surface density of galaxies
\citep{Feldmann2012,Narayanan2012}. We do not aim to provide yet
another scaling relation between $\alpha_{\rm CO}$ and galaxy
properties, merely to complement previous work. 

We present $\alpha_{\rm CO}$ of our modeled galaxies as a function of
cold gas metallicity in Figure \ref{fig:alpha_CO}. The predicted range
in  $\alpha_{\rm CO}$ is in good agreement with observations of the
CO-to-\h2 conversion factor
\citep{Bolatto2008,Leroy2011,Sandstrom2012}. We find good agreement with the predictions from the models of \citet{Narayanan2012}, but underpredict
$\alpha_{\rm CO}$ compared to the results of
\citet{Feldmann2012}. Our results
suggest that there is indeed a variation in the conversion between
\h2 mass and CO luminosity:  $\alpha_{\rm CO}$ rapidly decreases
  with metallicity at $Z/Z_\odot < 0.5$, driven by low CO
abundances. The variation of $\alpha_{\rm CO}$ with metallicity is
roughly flat for more metal-rich galaxies. This is in agreement with earlier theoretical work \citep{Feldmann2012,
  Narayanan2012} and observations \citep{Leroy2011}.

\subsection{HCN}
\label{subsec:HCN}
High-density star-forming regions are typically observed through
emission from HCN \citep[e.g.,][]{Loenen2008}. It is one of the most
abundant high-dipole-moment molecules that traces gas at densities of
the order $10^5-10^6$ cm$^{-3}$, several orders of magnitudes larger
than the low-excitation CO lines. HCN emission is thought to be
associated with star-forming GMC cores. \citet{Gao2004} found a tight
linear correlation between the FIR and HCN luminosity of galaxies,
which supports the scenario of HCN being linked to GMC cores.

We find that the HCN $J=1-0$ luminosity of galaxies increases monotonically with
FIR luminosity  (see Figure \ref{fig:scale_HCN}, left
panel). Assuming that the dust producing the FIR radiation is heated primarily
by young massive stars,
the FIR luminosity should be proportional to the instantaneous SFR, 
while the global SFR of a galaxy is
linearly proportional to the mass of dense molecular gas. Our predictions for the HCN $J=1-0$ luminosity of galaxies at $z=0.0$ are
slightly too low compared to observations. The HCN
$J=1-0$ luminosity in galaxies at redshift $z=1.2$ and $z=2.0$ also follows a
monotonically increasing trend with FIR luminosity, although there is
a minor offset to lower HCN luminosities with respect to our
predictions for local galaxies.  Unfortunately, the HCN luminosity of
high-redshift galaxies are not well constrained by
observations. However, our findings are in good agreement with the
upper limits presented in \citet{Gao2007}.  HCN $J=3-2$ traces the
densest regions of cold gas ($n \sim 10^7 \rm{cm}^{-3}$). Similar to
HCN $J=1-0$ our model slightly underpredicts the HCN $J=3-2$
luminosity of local galaxies.

\begin{figure}
\includegraphics[width = \hsize]{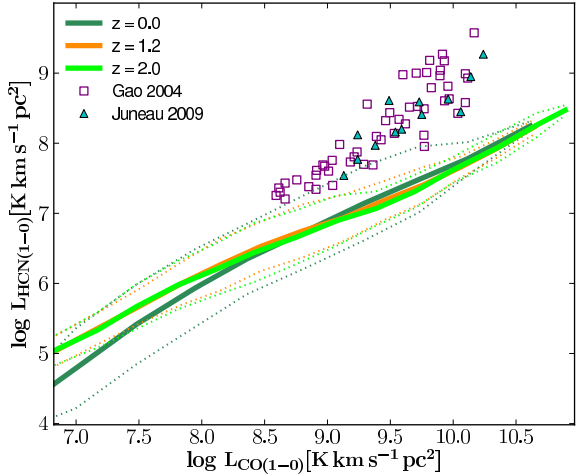}
\caption{The HCN $J=1-0$ luminosity of galaxies as a function of its
  CO $J=1-0$ luminosity at $z=0.0$,$z=1.2$, and $z=2.0$. Observations are
  taken from \citet{Gao2004} and
  \citet{Juneau2009}.\label{fig:HCN_CO}}
\end{figure}

The HCN/CO ratio of a galaxy is typically considered as a measure of
the very dense (few times $10^{5}$ cm$^{-3}$ to $10^{6}$ cm$^{-3}$)
versus less dense (few times $10^{3}$ to $10^4$ cm$^{-3}$) gas content
of a galaxy. In Figure \ref{fig:HCN_CO} we plot the HCN $J=1-0$
luminosity of a galaxy versus its CO $J=1-0$ luminosity. We
underpredict the HCN $J=1-0$ luminosity of galaxies as a function of
their CO $J=1-0$ luminosity compared with observations by
\citet{Gao2004} and \citet{Juneau2009} by approximately $0.1-0.2$
dex. This suggests that our modeled galaxies at $z=0.0$ have a
slightly too-low fraction of very dense cold gas. We find no
  significant difference between the galaxies at $z=2.0$ and $z=1.2$
  and our predictions for local galaxies.

\subsection{Neutral Carbon}
Because atomic carbon fine-structure emission proceeds through a
simple three-level system, detection of the two atomic carbon lines
[CI] (1-0) and [CI] (2-1) enables one to derive the excitation temperature
and neutral carbon mass of a galaxy independently of other
information. This is a powerful tool to study the properties of the
atomic gas in galaxies and to break some of the degeneracies
frequently found in CO studies \citep{Walter2011,Carilli2013}.

\begin{figure}
\includegraphics[width = \hsize]{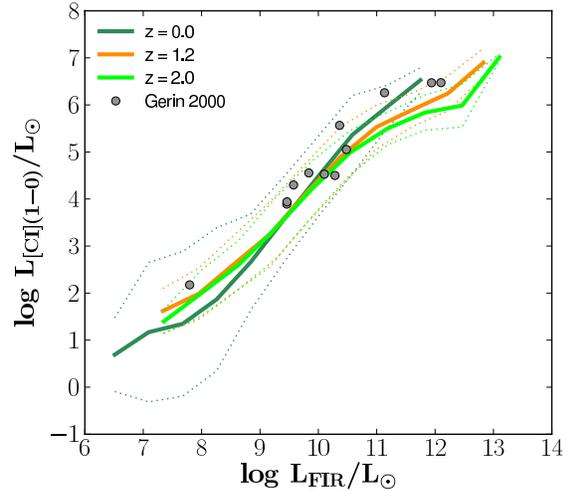}
\caption{[CI] (1-0) luminosity of
  modeled galaxies as a function of their FIR luminosity at redshifts
  $z=0.0$, $z=1.2$, and $z=2.0$. Observations are from
  \citet{Gerin2000}. \label{fig:scale_C}}
\end{figure}

We show the relation between FIR luminosity and [CI] (1-0) luminosity in
Figure \ref{fig:scale_C}, compared with observations taken from
\citet{Gerin2000}. The [CI] (1-0) luminosity increases monotonically with
FIR luminosity, in good agreement with observations over the entire
FIR luminosity range probed. Similar to CO and HCN,
we find that high-redshift galaxies have lower [CI] (1-0) luminosities
than their local counterparts (at $L_{\rm FIR} > 10^{10}\,\rm{L_\odot}$). This is driven by an
increased SFE, which effectively allows fewer carbon atoms to emit
radiation (see Section \ref{sec:discussion}).

\begin{figure}
\includegraphics[width = \hsize]{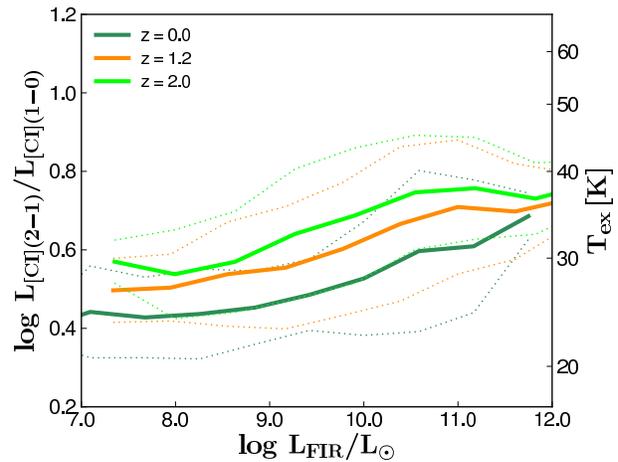}
\caption{The ratio between [CI] (2-1) and [CI] (1-0) (left axis) or carbon
  excitation temperature (right axis) as a function of FIR luminosity
  for modeled galaxies at $z=0.0$, $z=1.2$, and $z=2.0$. \label{fig:carbon_temp}}
\end{figure}

In local thermodynamic equilibrium the excitation temperature of
carbon can be derived via the formula
\begin{equation}
T_{\rm{ex}} = 38.8 \times \ln{\left(\frac{2.11}{R}\right)^{-1}}, 
\end{equation}
where $R=L_{\rm CI(2-1)}/L_{\rm CI(1-0)}$. We present the ratio $R$
and the excitation temperature of carbon of our modeled galaxies in
Figure \ref{fig:carbon_temp}. We find an increase in excitation
temperature at $z=0.0$ for galaxies with FIR luminosities brighter
than $L_{\rm FIR} > 10^9\,L_\odot$. Below this luminosity galaxies
have excitation temperatures of roughly $25\,\rm{K}$, after which they
increase to approximately $35\,\rm{K}$ for the brightest FIR objects. These excitation temperatures are in good agreement with temperatures
found in local galaxies ranging from $20 \,\rm{K}$ in our Galaxy to $50\,\rm{K}$ in starburst
environments \citep{Stutzki1997,Fixsen1999,Bayet2004}.
Modeled galaxies at high redshift are on average approximately $10$--$5\,
\mathrm{K}$ warmer than their local counterparts, and show a similar
increase in their temperature with FIR luminosity. The difference in
excitation temperature between galaxies at $z=1.2$ and $z=2.0$ is negligible.

The ratio between [CI] (1-0) and CO $J=1-0$ is often used to demonstrate
that atomic carbon can act as a good tracer of molecular gas in
external galaxies \citep{Gerin2000}. We find that this ratio is
roughly constant in our modeled galaxies at $z=0.0$ and high redshift with
[CI] (1-0)/CO $J=1-0$ $= 0.08$. This is in reasonable agreement with
observed ratios in the Milky Way \citep{Fixsen1999, Ojha2001,Oka2005},
the local Universe
\citep{Israel2001,Israel2002,Israel2003,Bayet2004,Israel2006} and at
high redshift \citep{Weiss2005,Walter2011}. Indeed, our models suggest
that carbon can act as a good tracer of the molecular mass of a
galaxy.

\begin{figure*}
\includegraphics[width = \hsize]{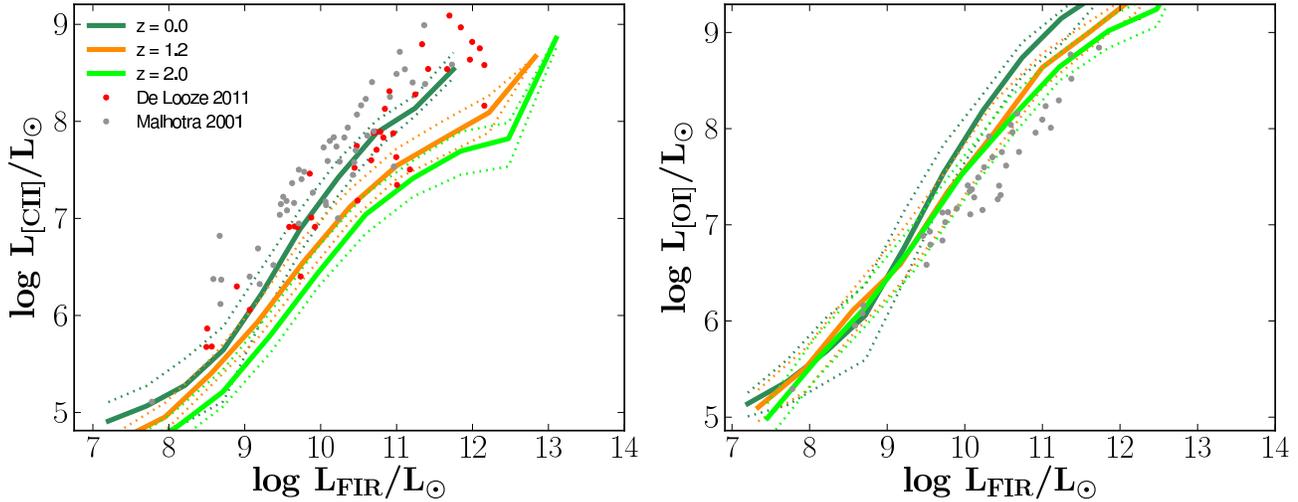}
\caption{Luminosity of the atomic cooling lines [CII] (158 $\mu$m) and
  [OI] (63 $\mu$m) as a function of their FIR luminosity, for galaxies
  at $z=0.0$, $z=1.2$, and $z=2.0$. Observations at $z=0.0$ are from
  \citet{Malhotra2001} and \citet{deLooze2011}.\label{fig:CII_OI_scale}}
\end{figure*}

\begin{figure*}
\includegraphics[width = \hsize]{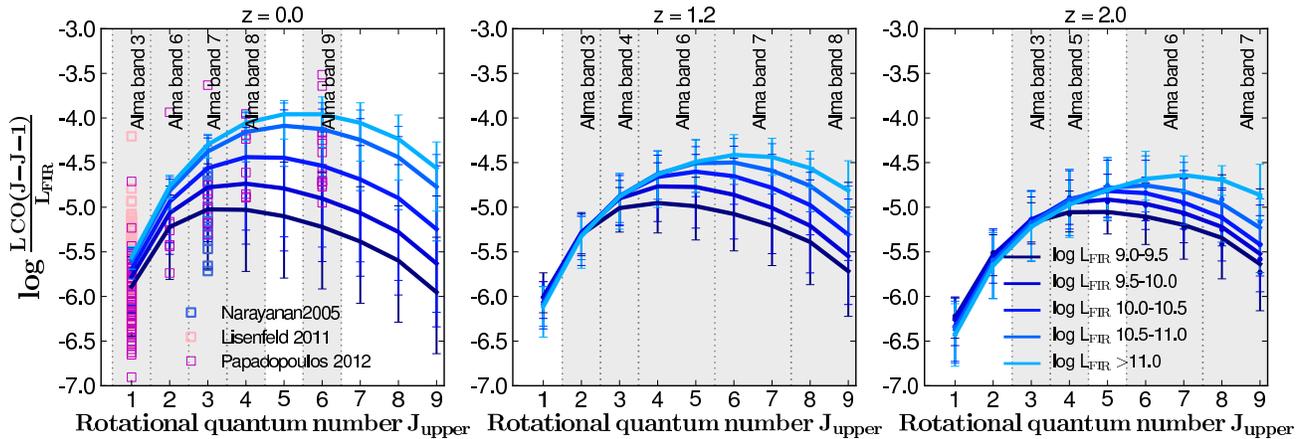}
\caption{CO $J= J - (J-1)$ to FIR luminosity ratio as a function of J
  level, separated in bins of FIR luminosity at redshifts $z=0.0$
  (left panel) and $z=1.2$ (middle panel) and $z=2.0$ (right panel). The CO $J= J - (J-1)$ to FIR
  luminosity ratio represents the cooling of the molecular gas
  through a CO line. Observations at $z=0.0$ are from
  \citet{Narayanan2005}, \citet{Lisenfeld2011}, and \citet{Papadopoulos2012}.\label{fig:CO_cooling_SLED}}
\end{figure*}

\begin{figure*}
\includegraphics[width = 0.9\hsize]{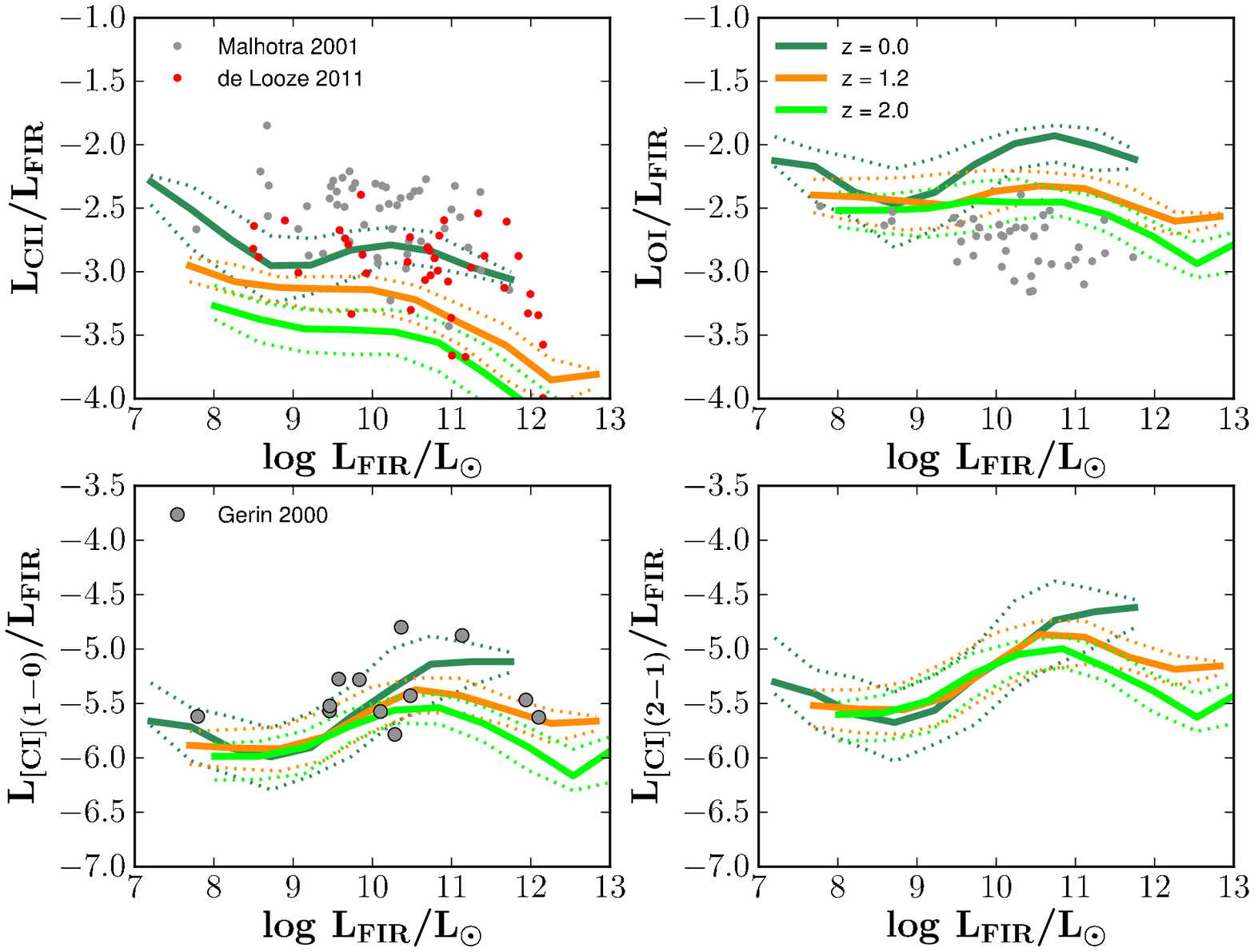}
\caption{Ratio of CI(1-0), CI(2-1), [CII] (158 $\mu$m), and [OI] (63
  $\mu$m) luminosity to FIR luminosity as a function of FIR
  luminosity at redshifts $z=0.0$, $z=1.2$, and $z=2.0$. Observations are
  from \citet{Gerin2000}, \citet{Malhotra2001}, and
  \citet{deLooze2011}. \label{fig:cooling_atoms}}
\end{figure*}

\subsection{[CII] (158 $\mu$m) and [OI] (63 $\mu$m) cooling lines}
[OI] (63$\mu$m) and [CII] (158$\mu$m) are the two dominant cooling
lines for interstellar gas. [CII] (158$\mu$m) is considered to
be a potentially promising indicator of a galaxy's SFR \citep[e.g.,][]{deLooze2011}. In
Figure \ref{fig:CII_OI_scale} (left panel) we present the [CII]
luminosity of our galaxies as a function of their FIR
luminosity. Observations are taken from \citet{Malhotra2001} and
\citet{deLooze2011}. We note that the authors apply different
approaches in estimating the total FIR luminosity. The
\citet{Malhotra2001} FIR estimates only cover a wavelength range from
40 to 500 $\mu$m, whereas our and the \citet{deLooze2011} FIR
luminosities cover a wider wavelength range. We therefore expect the
modeled galaxies to be more FIR luminous compared to the sample
presented in \citet{Malhotra2001}. We find decent agreement with the
observations over the entire range of FIR luminosities constrained by
observations. The [CII] luminosity of galaxies at $z=1.2$ and $z=2.0$ has a minor
offset towards lower values compared with the predicted luminosity of
local galaxies. This is again driven by a
higher SFE in galaxies at high redshift (see Sec. \ref{sec:discussion}).

Whereas [CII] (158$\mu$m) acts as an important coolant in the diffuse
medium, [OI] (63$\mu$m) cooling dominates in denser environments
with densities above a few times $10^3\, \rm{cm}^{-3}$. Our model predicts
an increase in [OI] (63$\mu$m) luminosity with FIR luminosity
(Figure \ref{fig:CII_OI_scale}, right panel). We overpredict the [OI]
luminosity of observed local galaxies, and the predicted slope is
steeper than suggested by observations \citep{Malhotra2001}.
The predicted [OI] luminosity in high-redshift galaxies are very
similar to the [OI] luminosities in local galaxies.

These results are encouraging, as they imply that we correctly model
and reproduce the [CII] (158$\mu$m) cooling line in local galaxies from more diffuse
regions. We look forward to being able to compare our results with luminosities of statistical
sets of typical star forming galaxies at high redshift. We discuss the contribution from [CII] and [OI] to the cooling of the
cold gas in the next section.

\subsection{Cooling}
In this subsection we discuss the dominant cooling processes acting in
the modeled galaxies. The net cooling rate and the dominant coolant of a
galaxy all depend heavily on the physical state of the cold gas. The
relative contribution of each coolant is therefore yet another
diagnostic to study the properties of the star forming cold gas.

We present the cooling contribution of each individual CO excitation
line (CO-to-FIR luminosity ratio) in the form of SLEDs in Figure
\ref{fig:CO_cooling_SLED}, separated into bins of FIR luminosity. Our
results are compared with observations taken from
\citet{Narayanan2005}, \citet{Lisenfeld2011}, and
\citet{Papadopoulos2012}. The more FIR-luminous a galaxy, the more
important the cooling through higher-J CO lines becomes. Cooling through CO $J=4-3$ and higher is especially dependent on the FIR
luminosity. This is tightly connected to the peaks in excitation level
as predicted by our model (see Figure \ref{fig:CO_SLED}), where we
also saw that more FIR-luminous galaxies peak at higher excitation
levels. These results suggest that in more luminous galaxies most of
the cooling takes place in denser regions, traced by CO $J=5-4$ and
higher.

Cooling through CO is approximately as efficient at high
  redshifts as it is in local galaxies for the low-FIR brightness
  galaxies in our sample ($L_{\rm FIR} <
  10^{10}\,L_\odot)$. Cooling through CO becomes less efficient with
  increasing redshift for the more FIR luminous galaxies in our
  sample. The dominant cooling occurs through higher CO J-states with increasing
redshift, independent of the FIR brightness of a galaxy. For example, at
$z=0.0$ the cooling curve for galaxies with a FIR
luminosity of $9.0 < \log{(L_{\rm FIR}/L_\odot)} < 9.5$ peaks at CO
$J=4-3$, whereas it peaks at CO $J=5-4$  and  $J=6-5$ in galaxies at
$z=1.2$ and $z=2.0$, respectively. Dominant CO cooling changes from
the $J=6-5$ state at $z=0.0$ to the $J=7-6$ state at $z=2.0$ for
galaxies with FIR brightnesses $10.0 < \log{(L_{\rm FIR}/L_\odot)} <
10.5$.  These results suggest that the cooling takes place in denser
regions, in good agreement with the CO SLED presented in Figure
\ref{fig:CO_SLED}.

We present the cooling rates of carbon, ionized carbon and oxygen in
Figure \ref{fig:cooling_atoms}. The cooling rate of all of these
chemical species is relatively constant with increasing FIR luminosity, both in
low- and in high-redshift galaxies. We find a small decrease in the
cooling rate of [CII] for galaxies brighter than $\log{(L_{\rm
    FIR}/L_\odot)} > 10$. This resembles the [CII]--FIR deficit in
galaxies, in which the [CII]/FIR ratio of galaxies goes down
with increasing FIR brightness \citep[see][and references therein for
a review]{Casey2014}; however, we do not see this trend for
galaxies with FIR luminosities fainter than
$10^{10}\,L_\odot$. At $z=1.2$ and $z=2.0$ the [CII]--FIR deficit is
prominent over the entire range in FIR luminosities probed. The contribution to the cooling budget from
oxygen, carbon and CO stays constant or even increases in this FIR luminosity
range. This suggests that the ionized carbon column saturates and
cooling takes place through other atomic and molecular species. We see that [CII] and [OI] are
the dominant coolants for the neutral gas, whereas the contribution to
the total cooling by neutral carbon is approximately two orders of
magnitude weaker at $z=0$. The contribution to the total
cooling by atomic carbon is even more negligible at $z=1.2$ and $z=2.0$.  The
negligible cooling from neutral carbon with respect to [CII] and [OI]
is not only supported by our compilation of observations in the local
Universe, but also by observations of SMGs at $z>2.0$
\citep{Walter2011}. CI (2-1) contributes slightly more to the total
cooling budget than CI (1-0), albeit still significantly less than
[CII] and [OI]. 

Adding up all the cooling rates we find that
carbon contributes less than one percent to the total cooling in the
modeled galaxies. Cooling through [CII] and [OI] dominates the total cooling
budget up to FIR luminosities of $10^{10.5}\,\rm{L}_\odot$. The
contribution to the total cooling budget by CO increases at higher
FIR luminosities, up to roughly 50 percent in the galaxies with
brightest FIR luminosities. This trend is independent of redshift.

\begin{figure}
\includegraphics[width = \hsize]{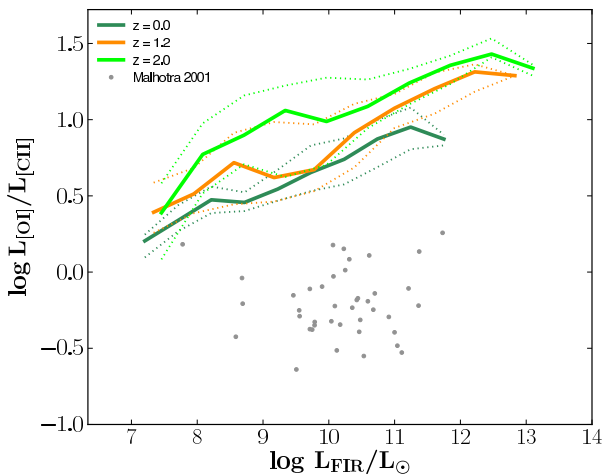}
\caption{The ratio between the [OI] (63 $\mu$m) and [CII] (158 $\mu$m)
  fine-structure lines as a function of FIR luminosity. This ratio
  traces the relative contribution to the cooling from dense
  ([OI], densities larger than $\sim 10^4\,\rm{cm}^{-3}$) and more
  diffuse ([CII], lower densities) regions. Observations
  are from \citet{Malhotra2001}. \label{fig:cooling_ratio}}
\end{figure}

Cooling through [CII] is approximately a dex lower
in galaxies at $z=1.2$ and $z=2.0$, whereas cooling through
[OI] decreases with only 0.5 dex. We present the ratio between [OI] and [CII] cooling in Figure
\ref{fig:cooling_ratio}. This ratio gives clear insight into which coolant becomes more dominant at the redshifts probed and as a
function of FIR luminosity.  We see an increase in [OI]/[CII]
with increasing FIR luminosity in galaxies at $z=0.0$, indicative of a
more dominant role of cooling through [OI]. The [OI]/[CII] ratio in
galaxies at high redshift is $\sim$ 0.5 dex higher than in local galaxies,
independent of FIR luminosity. A larger fraction of dense and warm
molecular clouds accounts for the more dominant contribution of
[OI] to the cooling.

\section{Discussion}
\label{sec:discussion}
In this paper we have presented a new set of theoretical models to
predict the sub-mm line emission properties of galaxies. Our approach
combines a semi-analytic galaxy formation model that explicitly tracks
the amount of gas in an atomic and molecular phase
\citep[Somerville, Popping \& Trager in prep.]{Popping2013} with a
fully three-dimensional radiative-transfer and non-local escape
probability line tracing code  \citep{Poelman2006,JP2011}. This approach provides sub-mm velocity maps, spectra and integrated luminosities
of mock galaxies. This model can provide detailed predictions of ISM
properties for statistical samples of galaxies that can be directly
compared with upcoming observations from ALMA, PdBI, LMT. Furthermore, it
drastically increases the number of predicted observables to be
constrained by current and future observations.
We did not include the effects from AGN and
shocks, which could add extra radiative or mechanical energy to the
system, heating up the gas. These processes do have the ability to
enhance the emission from high J-transition lines. Because we limit
ourselves to average galaxies on the star-forming ``main sequence'', we
do not expect these processes to be of significant importance for this
work.

We use our newly-developed model to study the ISM
properties in typical star forming galaxies at $z=2.0$, $z=1.2$, and $z=0$. We
aim to understand, from the model's perspective, if the physical state
of the gas in typical star forming galaxies during the
peak of the star formation history of the Universe is similar to
local counterparts or if the ISM physics that drives the SF is significantly
different.

We find that our approach is capable of reproducing sub-mm line luminosity
observations in both local and high-z galaxies. We correctly predict the
emission arising from CO, [CI], and [CII] at both low and
high redshift. These lines trace different regimes, ranging from
diffuse media with densities of the order $\sim
10^{3}-10^{4}\,\rm{cm}^{-3}$ to dense cores with densities of
$10^{6}\,\rm{cm}^{-3}$. Furthermore, they arise in both atomic and
molecular regions of the ISM. It is encouraging to see that our
approach is successful in reproducing these different states of the
ISM in galaxies. Unfortunately, to date, the sub-mm emission from
typical star-forming galaxies at high redshift is not well constrained
by observations. We believe that the predictive power of our model can
be valuable for future surveys with ALMA and other sub-mm telescopes,
but we also look forward to use
such surveys to better constrain our models.

We acknowledge that our models do not reach the necessary resolution
to self-consistently model the different phases of the ISM (which is
the case for all existing models of galaxy formation in a cosmological
context). We adopt simple assumptions about the distribution
of the ISM structure within a grid cell. Nevertheless, these simple
assumptions tell us something about the structure of the ISM. We
  find that  a lognormal density distribution for the cold gas within a galaxy is very appropriate to reproduce observations of the sub-mm emission in local
and distant galaxies covering a large range of densities probed.

Despite the success of our approach we underpredict the HCN luminosity of
galaxies at $z=0$ by a factor of three (see Figures \ref{fig:scale_HCN} and
\ref{fig:HCN_CO}). HCN has critical densities of $10^6$ cm$^{-3}$ and higher,
indicative that there are not sufficient molecular
clouds in the model galaxies that have the core densities required to reproduce the observations. A remedy would
require large gas reservoirs, a higher cold gas surface density, or a
larger fraction of
dense cores in the molecular clouds. The first two come with the risk of
predicting too high gas masses, SFRs and molecular fractions:
properties that are reproduced correctly by the current version of the SAM for local and high-redshift
observations \citep{Popping2013}. A larger number of dense cores would be
a more subtle solution but requires a slight revision of the log-normal
approach adopted in this work.

Our model does not correctly reproduce the slope and normalization of
  the observed relation between [OI] and FIR luminosity. The over-predicted [OI] luminosity could be due to
an over-estimate of the oxygen abundance by our models due to a not
yet well understood or implemented oxygen chemistry. This oxygen
chemistry should consider a larger fraction of the oxygen budget being depleted into
dust grains in the form of major oxygen bearing molecules, such as
H$_2$O and O$_2$. This is the same problem as found between models and
observations of Galactic molecular clouds, where the observed
abundance of H$_2$O and O$_2$ molecules is at least one order of
magnitude lower than expected from current chemical models
\citep[e.g.,][]{vanDishoeck1993, Spaans2001, Goldsmith2011, Liseau2012}. Hence, we
should consider, in a future work, a scenario where a significant
fraction of the oxygen budget is lost into icy mantles of dust grains
in high column density ($\rm{N}(\rm{H}_2)>10^{22}\,\rm{cm}^{-2})$ clouds.

At fixed redshift the CO SLED of FIR-luminous galaxies peaks at higher
excitation levels than in less FIR-luminous galaxies (Figure
\ref{fig:CO_SLED}). Similarly, CO line ratios increase as a function
of the SFR in galaxies and the cooling through molecular lines occurs in
higher-J CO levels in more FIR bright galaxies. These trends may
partially be driven by a low CO abundance in the galaxies with lowest
metallicities. Above FIR luminosities of $10^{9
  -9.5}\,\rm{L}_\odot$ the cold gas in galaxies is equally enriched in CO. Furthermore the [OI]/[CII] cooling-line ratio increases with FIR
luminosity. These results indicate that FIR-luminous galaxies are being built up by
warmer and denser molecular clouds than less-bright galaxies. 

We briefly studied the behavior of the CO-to-\h2 conversion factor
$\alpha_{\rm CO}$. We find good agreement with observations of
$\alpha_{\rm CO}$ for galaxies over a large range of
  metallicities.  We predict a steep decline in $\alpha_{\rm CO}$ at the lowest
metallicities $Z' < 0.1$ and a flat distribution of $\alpha_{\rm CO}$
at higher metallicities. These results are consistent with other theoretical results
\citep{Feldmann2012,Narayanan2012} and are mostly driven by the low CO
abundance of low-metallicity gas. The approach presented in this work
is still somewhat over-simplified. A detailed study of $\alpha_{\rm CO}$ should
include a proper chemical network, simultaneously solving for the  CO
abundance and \h2 abundance of cold gas as a function of density,
optical depth, impinging radiation field and individual elemental abundances. This is a computationally expensive exercise, and beyond the scope of this work.

The carbon excitation temperature in galaxies at $z=1.2$ and $z=2.0$ is warmer (approximately $10 \,\mathrm{K}$) than in galaxies at $z=0$. This
is suggestive of a warmer atomic ISM in star forming galaxies at
high redshift. Our predicted atomic ISM temperatures are in good
agreement with results from \citet{Weiss2005} and \citet{Walter2011},
who found that the carbon excitation temperature in SMGs at $z>2.0$ is
around $30 \,\mathrm{K}$. Furthermore, the carbon-to-CO ratio also is constant with
redshift, demonstrating that the carbon and CO $J=1-0$ emission arise
in roughly the same medium. Neutral carbon provides only a negligible
contribution to the total cooling in a galaxy.

We have compared the sub-mm line properties of modeled galaxies at
$z=0.0$, $z=1.2$, and $z=2.0$ in some detail, with the aim of understanding
whether the gas properties of galaxies at high redshift are different from
local counterparts. Within the models, we find multiple examples of
evidence that suggest that SF in galaxies during the SF activity
peak of the Universe takes place in
much denser and warmer environments than SF in similar galaxies at
$z=0$. Galaxies at high redshift show a peak in their CO SLED at
higher CO J-states (Figure \ref{fig:CO_SLED}), have higher CO
$J=5-4$/CO $J=2-1$ and CO $J=7-6$/CO $J=2-1$ ratios as a function of
several global galaxy properties (Figure \ref{fig:CO_ratio}). Furthermore, cooling
through the [OI] 63 $\mu$m fine structure line becomes more dominant
and molecular cooling occurs through higher J CO levels. All these
results are indicative of a difference in gas density and temperature
between typical star forming galaxies in the local Universe and at
high redshift. Rather than being in a similar physical state as local
galaxies, the cold gas in typical star forming galaxies at high redshift
appears to have different physical properties. We find that galaxies
at $z=1.2$ have lower excitation temperatures and densities
than similar galaxies at $z=2.0$. This indicates that different ISM
conditions are already visible right after the actual SF peak of the
Universe ($z<2.0$).

The CO line ratios of galaxies remain constant with time when plotted
as a function of \h2 surface density. Although the shape of the
CO SLED is determined by a number of quantities such as density,
turbulence and temperature, this suggests that we can describe the CO
SLED to first order as a function of only the \h2 surface density. One should keep in
mind that within our model the SFR of a galaxy (which will ultimately
set the cold gas temperature) is a function of \h2 surface density as
well. We therefore believe our results are also in good agreement with
the work by \citet{Narayanan2014}, who found that the CO
SLED of a galaxy is well correlated with a galaxy's SFR surface density.

Our results provide an explanation for the shorter depletion times
observed in galaxies at high redshift \citep{Genzel2010}. The higher
densities will shorten the free-fall time of the molecular clouds, and
thus enhance the rate at which gas may condense and form stars. This
increment in SFE provides an attractive way to increase the SFR
density of the Universe.  A denser medium allows less gas-rich
galaxies to contribute to the total SFR more efficiently than less
dense objects with similar SFR, but driven by lower density (and more)
gas. This behavior is reflected in our models through the lower
luminosities from low J-states of CO, HCN, [CI] and [CII] in galaxies at
$z=1.2$ and $z=2.0$ compared to local galaxies.  There are effectively fewer atoms
and molecules that can add to the galaxy's emission for the same net
amount of FIR radiation. The available observations for high-redshift
galaxies do not convincingly support such a trend
\citep{Tacconi2010,Tacconi2013}. It is currently premature to draw a
firm conclusion on this point, given the small sample size and strong
bias towards extreme star formation of the present observations.

Observationally, the difference in ISM properties between local and
high-redshift galaxies is not well
constrained. \citet{Dannerbauer2009} suggests that the star-forming
conditions in a typical star forming galaxy are similar to local
galaxies. These observations, however, only go out to CO $J=3-2$ and
do not probe the CO excitation levels where our models show a clear
difference between local and high-redshift galaxies. Furthermore, their
conclusion is based on only two galaxies. We hope that future
observations of (parts of) the CO SLED with ALMA will better constrain
the gas properties of typical star-forming galaxies at high redshifts.

A first attempt at combining semi-analytic models of galaxy formation
with radiative-transfer codes was presented in \citet{Lagos2012}. The
authors parameterize modeled galaxies with a single cold gas density,
UV radiation field, metallicity and X-ray intensity and use a library
of pre-calculated radiative-transfer models to infer conversion
factors between \h2 mass and CO line-intensities. The authors find
similar trends with FIR luminosity for the luminosity from low
excitation CO lines as we do. We predict higher luminosities and a
larger contribution to the molecular cooling from high J CO lines ($J
= 5-4$ and up) compared with their results. We ascribe this difference
to their representation of a galaxy with a single non-variable ISM
density of $10^4\,\rm{cm}^{-3}$ (although see their section 5 for a
discussion about using different densities). In our approach galaxies are
represented by the summation of multiple molecular clouds with varying
density. This becomes especially relevant for CO lines like $J=5-4$
and up, as the critical densities of these lines lie almost an order
of magnitude above $10^4\,\rm{cm}^{-3}$, if not
more. \citet{Lagos2012} only study the CO luminosities of their
modeled galaxies, and do not present predictions for emission from
other atomic and molecular species. We have shown that the ability to examine
more atomic and molecular species provides a more
detailed picture of the ISM and its different phases.

\section{Summary}
\label{sec:summary}
In this paper we developed new models of the sub-mm line emission
from atomic and molecular species for statistical sets of
galaxies. We summarize our main findings below:
\begin{itemize}
\item We successfully reproduce observed scaling relations for the line luminosities emitted by CO, C,
  C$^+$, and O for nearby galaxies. These atomic and molecular species trace a
  wide range in molecular cloud properties (gas temperatures
  and densities), indicating that our model correctly
  reproduces the multi-phase structure of molecular clouds.

\item The peak excitation level of the CO SLED in modeled galaxies
  increases with FIR luminosity, as well as with redshift. 

\item CO line ratios of our modeled galaxies show a clear increasing
  trend with global SF tracing properties (SFR, FIR luminosity, cold gas surface density)
  and with redshift (SFR, FIR luminosity). Most notable changes in CO line ratios are
  achieved when using CO $J=5-4$/CO $J=2-1$ and higher. These are lines
  that ALMA can potentially observe at $z > 1.0$. The CO line ratios of galaxies are well correlated with galaxy
  \h2 surface density independent of redshift.

\item Galaxies at $z=1.2$ and $z=2.0$ have weaker low-J line
  luminosities of CO, HCN, and [CI]  than galaxies at $z=0.0$ with
  similar SFR and FIR luminosity.

\item The atomic gas properties of galaxies, as traced through neutral carbon,
  in local galaxies and galaxies at high redshift are similar, with no notable
  difference in carbon excitation temperature.

\item The [OI]/[CII] cooling line ratio is higher in
  galaxies at high redshift than in local galaxies, suggesting that cooling
  predominantly takes place in denser regions.

\item Our model results indicate that SF in galaxies at high redshift takes place in denser and warmer environments than in local counterparts. This
  suggests that SF during the peak of the SF activity in the Universe
  is not driven by gas in a similar physical state as in local galaxies, but by
  significantly different ISM conditions. Galaxies belonging to the
  tail of the SF activity peak of the Universe ($z=1.2$) are already
  less dense and cooler than counterparts during the actual peak of SF
  activity ($z=2.0$).
\end{itemize}

Observations with the current and next generation of sub-mm telescopes such as ALMA,
LMT, and PdBI will allow us to test the predictions made in
this work for large samples of galaxies. Not only can these
observations be placed in their proper physical context using models as
presented in this work, they will also further constrain models of
galaxy formation and evolution. Most notable results can be obtained
by observing line ratios of CO in high- and low-J states (tracing a
critical density difference of at least an order of magnitude). Not
only will observations of tracers of such different gas phases give
insight into the dominant gas distribution, they also provide strong
constraints on the CO SLED from which density and excitation
temperatures can be derived. Besides CO emission, atomic cooling lines
[OI] 63 $\mu$m and [CII] 158 $\mu$m also provide great potential to
study ISM properties in galaxies. We predict a difference in
[OI]/[CII] ratios with redshift, directly related to the density and
temperature of the ISM.

Our approach, in combination with future surveys, holds great
potential for further understanding galaxy formation. It allows us to
constrain not only the amount of cold gas in galaxies, but also the
actual density and temperature distribution of the gas that will set
the rate of formation of stars and the build-up of stellar
discs. Any physical process that acts on the cold gas content of a galaxy
should result in a temperature and density distribution and line
emission along the atomic and/or molecular energy ladder in close agreement with observations.
Comparing sub-mm observations across cosmic time with our
model therefore provides an attractive way to constrain the physics acting on the
cold gas that account for the gas content and SF activity in galaxies
during the peak of SF in the Universe and its turnover at lower redshifts. [CII] luminosities
will be a powerful way of constraining model predicted SFRs in the high-redshift Universe. Furthermore, when
including the spatial and velocity information, we are in the position
to make predictions for observed CO disc sizes and velocities. This
offers the unique possibility to address the topic of disc angular
momentum at redshifts above $z=0$.

\section*{Acknowledgments}
We thank Andrew Baker, Daniella Calzetti, Colin Norman, Desika
Narayanan, Padelis Papadopoulos and Eve Ostriker for stimulating discussions, and Linda Tacconi for providing observational
data. We thank the referee, Guinevere Kauffmann, for suggestions that have improved
  this paper. GP acknowledges NOVA (Nederlandse Onderzoekschool voor
Astronomie) and LKBF (Leids Kerkhoven-Bosscha Fonds) for funding.

\bibliographystyle{mn2e_fix}
\bibliography{references}

\end{document}